\documentclass[11pt]{amsart}
\reversemarginpar
\newcommand{\commentout}[1]{}

\usepackage{amsmath}
\usepackage{amssymb}

\newcommand{\nwc}{\newcommand}

\newcommand{\invf}{\cF^{-1}_2}
\newcommand{\myphi}{\Phi_{(\eta,\rho)}}
\newcommand{\minrg}{|\min{(\rho,\gamma^{-1})}|}

\newcommand{\xvec}{\vec{\mathbf x}}
\newcommand{\kvec}{{\vec{\mathbf k}}}
\newcommand{\lt}{\left}
\newcommand{\rt}{\right}
\newcommand{\vas}{\varepsilon}
\newcommand{\lan}{\left\langle}
\newcommand{\ran}{\right\rangle}
\newcommand{\tvas}{W_z^\vas}
\newcommand{\psiep}{W_z^\vas}
\newcommand{\wep}{W^\vas}

\newcommand{\wepz}{W_z^\vas}

\newcommand{\veptil}{\tilde{\ml L}_z^\vas}

\newcommand{\cv}{{\ml L}^\ep_z}
\newcommand{\cvtil}{\tilde{{\ml L}}^\ep_z}

\newcommand{\ks}{\tilde{k}}
\newcommand{\bx}{\mathbf x}
\newcommand{\bu}{\mathbf u}
\newcommand{\bD}{\mathbf D}
\newcommand{\bp}{\mathbf p}
\newcommand{\bq}{\mathbf q}
\newcommand{\by}{\mathbf y}

\newcommand{\pdgx}{\bp\cdot\nabla_\bx}

\newcommand{\corr}{r_{\eta,\rho}}
\newcommand{\rinf}{r_{\eta,\infty}}

\nwc{\nwt}{\newtheorem}

\nwt{prop}{Proposition}
\nwt{proposition}{Proposition}
\nwt{lemma}{Lemma}
\nwt{assumption}{Assumption}
\nwt{theorem}{Theorem}
\nwt{corollary}{Corollary}
\nwt{remark}{Remark}
\nwt{definition}{Definition} 

\nwc{\bal}{\begin{align}}
\nwc{\be}{\begin{equation}}
\nwc{\ben}{\begin{equation*}}
\nwc{\bea}{\begin{eqnarray}}
\nwc{\beq}{\begin{eqnarray}}
\nwc{\bean}{\begin{eqnarray*}}
\nwc{\beqn}{\begin{eqnarray*}}
\nwc{\beqast}{\begin{eqnarray*}}

\nwc{\eal}{\end{align}}
\nwc{\ee}{\end{equation}}
\nwc{\een}{\end{equation*}}
\nwc{\eea}{\end{eqnarray}}
\nwc{\eeq}{\end{eqnarray}}
\nwc{\eean}{\end{eqnarray*}}
\nwc{\eeqn}{\end{eqnarray*}}
\nwc{\eeqast}{\end{eqnarray*}}

\nwc{\ep}{\varepsilon}
\nwc{\eps}{\varepsilon}
\nwc{\ept}{\epsilon}
\nwc{\vrho}{\varrho}
\nwc{\orho}{\bar\varrho}
\nwc{\ou}{\bar u}
\nwc{\vpsi}{\varpsi}
\nwc{\lamb}{\lambda}

\nwc{\nn}{\nonumber}
\nwc{\bm}{\boldmath}
\nwc{\mf}{\mathbf}
\nwc{\mb}{\mathbf}
\nwc{\ml}{\mathcal}

\nwc{\IA}{\mathbb{A}} 
\nwc{\IB}{\mathbb{B}}
\nwc{\IC}{\mathbb{C}} 
\nwc{\ID}{\mathbb{D}} 
\nwc{\IM}{\mathbb{M}} 
\nwc{\IP}{\mathbb{P}} 
\nwc{\II}{\mathbb{I}} 
\nwc{\IE}{\mathbb{E}} 
\nwc{\IF}{\mathbb{F}} 
\nwc{\IG}{\mathbb{G}} 
\nwc{\IN}{\mathbb{N}} 
\nwc{\IQ}{\mathbb{Q}} 
\nwc{\IR}{\mathbb{R}} 
\nwc{\IT}{\mathbb{T}} 
\nwc{\IZ}{\mathbb{Z}} 

\nwc{\cE}{{\ml E}}
\nwc{\cP}{{\ml P}}
\nwc{\cQ}{{\ml Q}}
\nwc{\cL}{{\ml L}}
\nwc{\cR}{{\ml R}}
\nwc{\cV}{{\ml L}}
\nwc{\cT}{{\ml T}}
\nwc{\crV}{{\ml L}_{(\delta,\rho)}}
\nwc{\cC}{{\ml C}}
\nwc{\cA}{{\ml A}}
\nwc{\cK}{{\ml K}}
\nwc{\cB}{{\ml B}}
\nwc{\cD}{{\ml D}}
\nwc{\cF}{{\ml F}}
\nwc{\cS}{{\ml S}}
\nwc{\cM}{{\ml M}}
\nwc{\cG}{{\ml G}}
\nwc{\cH}{{\ml H}}
\nwc{\bk}{{\mb k}}
\nwc{\cbz}{\overline{\cB}_z}

\nwc{\pft}{\cF^{-1}_2}

\begin{document}

\title{White-Noise and Geometrical Optics Limits of Wigner-Moyal 
Equation for
Wave Beams 
in Turbulent Media}

\author{Albert C. Fannjiang
 }
\thanks{Department of Mathematics,
University of California at Davis,
Davis, CA 95616
Internet: fannjian@math.ucdavis.edu
The research is supported in part by The Centennial Fellowship
from American Mathematical Society,
the UC Davis Chancellor's Fellowship
and U.S. National Science Foundation grant DMS 0306659.
}

\begin{abstract}
Starting with the Wigner distribution formulation for  beam
wave propagation in H\"{o}lder continuous
non-Gaussian random refractive index fields
we show that the wave beam regime naturally leads to the
white-noise scaling limit and converges to a Gaussian
white-noise model which is characterized  by the
martingale problem associated to a stochastic
differential-integral equation of the It\^o type. In the 
simultaneous 
geometrical optics  the convergence to
the Gaussian white-noise model for 
the Liouville equation is also established 
if the ultraviolet cutoff  or the Fresnel number vanishes
sufficiently slowly. The advantage of the
Gaussian white-noise model is that its $n$-point
correlation functions are  governed by  closed
form equations.
\end{abstract}

\maketitle

\section{Introduction}

Laser beam propagation in the turbulent atmosphere is governed  by
the  classical wave equation with  a randomly inhomogeneous refractive index field
\[
n(z, \bx)=\bar{n}(1+\tilde{n}(z, \bx)),\quad (z,\bx)\in \IR^3
\]
where $\bar{n}$ is the mean and $\tilde{n}(\bx)$ is
the fluctuation of the refractive index field. 
We seek the solution of  the form
$E(t,z,\bx)=\Psi(z,\bx)\exp{[i\bar{n}(kz-\omega t)]}
+\hbox{c.c.}$ where
$E$ is   the (scalar) electric field, 
$k$ and $w=k c_0/\bar{n}$ are the carrier wavenumber and frequency,
respectively, 
with $c_0$ being the wave speed in
vacuum.  Here and below  $z$ and $\bx$ denote the variables in the longitudinal
and transverse directions  of the wave beam, respectively.

In the  forward scattering approximation \cite{St},  the modulation
$\Psi$ is approximated by the solution of  the parabolic wave equation which
after nondimensionalization with respect to some reference lengths $L_z$ and
$L_x$  in the
longitudinal and transverse directions,
respectively, has this form
\beq
i\ks \frac{\partial \Psi}{\partial z} +
\frac{\gamma}{2}\Delta\Psi+
 \ks^2 {k_0 L_z} \tilde{n}(z L_z,
{\bx L_x})\Psi=0,\quad\Psi(0,\bx)=\Psi_0(\bx) \in L^2(\IR^d),\quad d=2
\label{0.3}
\eeq
where 
$\ks=k/k_0$ is 
the normalized wavenumber with respect to 
the central wavenumber $k_0$ and $\gamma$ is
the Fresnel number 
\[
\gamma=\frac{L_z}{k_0 L_x^2}.
\]

A widely used model for the fluctuating refractive index field $\tilde{n}$
is a spatially homogeneous random field (usually assumed to be
Gaussian) with  the spatial
structure function
\[
D_n(|\xvec|)=\IE[\tilde{n}(\xvec+\cdot)-\tilde{n}
(\cdot)]^2= C_n^2 |\xvec|^{2/3},\quad
|\xvec|\in (\ell_0, L_0), \quad \xvec=(z,\bx)\in \IR^{d+1},\quad d=2
\]
where $\ell_0$ and $L_0$ are the inner and outer scales, respectively.
Here and below $\IE$ stands for ensemble average.

The refractive index structure function has a spectral
representation
\beq
\label{sf}
D_n(|\xvec|)=8\pi \int^\infty_0\Phi_n(|\kvec|)\lt[
1-\frac{\sin{(|\kvec||\xvec|)}}{|\kvec||\xvec|}\rt] |\kvec|^2d|\kvec|,
\quad\kvec\in
\IR^{d+1}
\eeq
with the Kolmogorov spectral density
\beq
\label{3.2}
\Phi_n(|\kvec|)=0.033C_n^2|\kvec|^{-11/3}, \quad|\kvec|\in (\ell_0,
L_0).
\eeq
Here the structure parameter $C_n^2$ depends in
general
on the temperature gradient on the scales larger than
$L_0$. See, e.g., \cite{MY}, \cite{Hi}
and \cite{BO} for
more sophisticated models of turbulent refractive index fields.

In this paper we will
consider a general class of  spectral
density parametrized by $H\in (0,1)$
and satisfying the upper bound 
\beq
\label{p2}
\Phi( \kvec)
&\leq&K
(L_0^{-2}+|\kvec|^2)^{-H-1/2-d/2}\lt(1+\ell^{-2}_0|\bk|^2\rt)^{-2},\quad
\kvec=(\xi,\bk)\in
\IR^{d+1},\quad d=2\label{powerH}
\eeq
for some positive constant $K<\infty$.
$L_0$ and $\ell_0$ in (\ref{p2}) are
the infrared and ultraviolet cutoffs.
The ultraviolet cutoff is physically
due to dissipation on the small scales which normally results in
a Gaussian decay factor \cite{MY}. 
We
are particularly interested in the regime where the ratio  $L_0/\ell_0 $ is
large  as in the high Reynolds number turbulent atmosphere.

Let us introduce the non-dimensional parameters that are pertinent to our
scaling:
\[
\ep = \sqrt{\frac{L_x}{L_z}},\quad
\eta=\frac{L_x}{L_0},\quad
\rho=\frac{L_x}{\ell_0}.
\]
In terms of the parameters and the power-law spectrum in
(\ref{powerH}) we  rewrite (\ref{0.3}) as
\beq
i\ks \frac{\partial \Psi^\ep}{\partial z} +
\frac{\gamma}{2}\Delta\Psi^\ep+\frac{\ks^2}{\gamma}\frac{\mu}{\ep}
V(\frac{z}{\ep^2},
\bx)\Psi^\ep=0, \quad\Psi^\ep(0,\bx)=\Psi_0(\bx)
\label{para}
\eeq
with
\beq
\label{inhom}
\mu = 
\frac{\sigma L_x^{H}}{\ep^3} 
\eeq
where $\sigma$ the  standard variation of 
the homogeneous field $\tilde{n}(z, \bx)$
and
$V$ is the normalized refractive index field
with a spectral density satisfying the
upper  bound
\beq
\label{power}
\hspace{1cm} \Phi_{\eta,\rho}( \kvec)\leq K
(\eta^2+|\kvec|^2)^{-H-1/2-d/2}\lt(1+\rho^{-2}|\bk|^2\rt)^{-2},\quad
\kvec\in
\IR^{d+1}, \quad H\in (0,1)
\eeq
for some positive constant $K$.


The generalized von K\'arm\'an spectral density \cite{FGV},
\cite{MY}
\beq
\label{vk}
\Phi_{vk}(\kvec)=2^{H-1}\Gamma(H+\frac{d+1}{2})
\eta^{2H} \pi^{-(d+1)/2}(\eta^2+|\kvec|^2)^{-H-1/2-d/2}
\eeq
corresponds to the isotropic covariance function
\[
B_{vk}(\xvec)=\IE\lt[V(\xvec+\cdot)
V(\cdot)\rt]=|\eta\xvec|^H K_H(\eta|
\xvec|),\quad
\xvec=(z,\bx)\in
\IR^{d+1}
\]
where $K_H$ is a Bessel function of the third kind given by
\[
K_H(z)=\int^\infty_0 \exp{\lt[-z \frac{e^t+e^{-t}}{2}\rt]}
\frac{e^{Ht}+e^{-Ht}}{2} dt.
\]
For $H=1/2$ we have the exponential covariance function
$B_{vk}(\xvec)=\exp{[-\eta |\xvec|]}$. The additional
ultraviolet cutoff imposed
in the upper bound (\ref{power}) would then 
 give rises to 
the covariance function
\[
B(\xvec)=G\star B_{vk}(\xvec)
\]
where $G$ is the inverse Fourier transform of the
cutoffs.

For high
Reynolds number one has
$L_0/\ell_0=\rho/\eta\gg 1$ and thus a wide range of scales in the
power spectrum (\ref{power}).
Note that in the worst case scenario the refractive index field
loses spatial differentiability  as ${\rho}\to\infty$  and
homogeneity
 as $\eta\to 0$. 
 The Gaussian field
 with its spectral density given by the right side
 of (\ref{power}) has  $H$ as the
  upper limit of the H\"older exponent of the sample field.
 The Kolmogorov spectrum has the
 exponent
 $H=1/3$.
 Since our result does not depend on $d$ we hereafter
 take it to be any positive integer.
 
 Although we do not assume isotropic spectral densities,
 the spectral density always satisfies the basic symmetry:
 \beq
 \label{sym}
 \Phi_{(\eta,\rho)}(\xi,\bk)=\Phi_{(\eta,\rho)}(-\xi, \bk)
 =\Phi_{(\eta,\rho)}(\xi,-\bk),\quad \forall (\xi, \bk)\in \IR^{d+1}.
 \eeq
 In other words, the spectral density is invariant under change
 of sign in any component of the argument because
 it is a characteristic function of a real-valued  stationary process.

We also assume  that $V_z(\bx)\equiv V(z,\bx)$ is a
square-integrable,
$z$-stationary and
$\bx$-homogeneous process with the (partial) spectral
representation 
\beq
\label{parts}
V_z(\bx)=\int \exp{(i\bp\cdot \bx)}
\widehat{V}_z(d\bp)
\eeq
 where the process $\widehat{ V}_z(d\bp)$ is the
$z-$stationary orthogonal spectral measure satisfying
\beq
\label{partp}
\IE\lt[ \hat{V}_z( d\bp)\hat{V}_z(d\bq)\rt]
=\delta(\bp+\bq)\lt[\int \Phi( w, \bp)dw \rt]d\bp d\bq.
\eeq
We do
{\em not} assume  the Gaussian  property
but instead a quasi-Gaussian property
(see Assumption~1, 2 and 3 in
Section~\ref{assumption} for precise statements).

If the observation scales $L_z$ and $L_x$ are the longitudinal and
transverse scales, respectively, of the wave beam then
$\ep\ll 1$ corresponds to  a long, narrow  wave beam. The white-noise scaling then
corresponds to
$\ep\to 0$ with a fixed
$\mu$. For convenience  we set $\mu=1$.
The  white-noise scaling limit $\ep\to 0$ of Eq. (\ref{para}) is analyzed
in \cite{para} (see also \cite{BCF}). The limit $\gamma\to 0$ corresponds to
the geometrical optics limit.
In this paper we study the higher moments behavior
in both white-noise and geometrical optics limits
by considering the Wigner transform of the modulation function.

Our method is also suitable for the situation where
 deterministic large-scale  inhomogeneities are present. One type
of slowly varying,  large-scale inhomogeneities is multiplicative and
can be modeled by a bounded smooth deterministic function $\mu=\mu(z,\bx)$
 due to variability
of any one of the three factors in (\ref{inhom}) (see, e.g.,
\cite{BO}, \cite{An} for models with slowly varying
$\sigma$).
The second type is additive and can be modeled
by adding to $\ep^{-1}\mu V(z\ep^{-2},\bx)$ a smooth 
background $V_0(z,\bx)$. Altogether we can treat
the random refractive index field of the general type
\[
V_0(z,\bx)+\frac{\mu(z,\bx)}{\ep}
V(\frac{z}{\ep^2},\bx)
\]
with a bounded smooth deterministic modulation and background in the parabolic
wave equation (\ref{para}).
 We describe the results
in Section~\ref{inhom2} but omit
the details of the argument for  simplicity
of presentation. As  the small-scale turbulent fluctuations
are invariably embedded in a structure 
determined by large-scale geophysics  this generalization is
important for the practical application of the
scaling limits.

\subsection{Wigner distribution and Wigner-Moyal equation}
The Wigner transform of $\Psi^\ep$, called
the Wigner distribution, is defined as
\beq
\label{0.11}
W^\eps_z(\bx,\bp)=\frac{1}{(2\pi)^d}\int
e^{-i\bp\cdot\by}
\Psi^\eps (z,\bx+\frac
{\gamma\by}2){\Psi^{\eps*}(z,\bx-\frac{\gamma\by}{2})}d\by.
\eeq
One has the following bounds from (\ref{0.11})
\[
\|W^\ep_z\|_\infty\leq
(2\gamma\pi)^{-d}\|\Psi^\ep(z,\cdot)\|_2^2,\quad
\|W^\ep_z\|_2=(2\gamma\pi)^{-d/2}\|\Psi^\ep(z,\cdot)\|_2^2
\]
\cite{Fo}, \cite{GMM}, \cite{LP}.
The Wigner distribution 
has many important properties.
For instance,
it is real and its $\bp$-integral is the modulus square of the function $\phi$,
\bea\label{2.2.2}
\int _{\IR^{d}}W^\ep(\bx,\bp)d\bp=|\Psi^\ep(\bx)|^2 ,
\eea
so we may think of $W(\bx,\bp)$ as wave number-resolved mass density.
Additionally, its $\bx$-integral is
\[
\int_{\IR^d}W^\ep(\bx,\bp)d\bx=(\frac{2\pi}{\gamma})^{d}
|\widehat\Psi^\ep|^2(\bp/\gamma).
\]
The energy flux
is expressed through $W^\ep(\bx,\bp)$ as
\bea\label{2.2.3}
\frac 1{2i}(\Psi\nabla\Psi^*-\Psi^*\nabla\Psi)=
\int_{\IR^d}\bp W^\ep(\bx,\bp)d\bp
\eea
and its second moment in $\bp$ is
\bea\label{2.2.4}
\int |\bp|^2W(\bx,\bp)d\bp  =|\nabla\Psi^\ep (\bx)|^2.
\eea
In view of these properties it is
tempting to think of the Wigner distribution as
a phase-space probability density, which is unfortunately not
the case, since it is not everywhere non-negative.
Nevertheless, the Wigner distribution is a useful tool for analyzing
the evolution of wave energy in  the phase space. Moreover, in the
recent development of time reversal of waves in which
a part of the waves is received, phase-conjugated and
then back-propagated toward the source the refocused 
wave field is given by a Wigner distribution of  mixed-state type
(see (\ref{mix}) below) \cite{BPZ}, \cite{PRS}, \cite{tire}.

Moreover, the Wigner distribution, written as $\wepz(\bx,\bp)=
\wep(z,\bx,\bp)$, satisfies an evolution equation,
called the Wigner-Moyal equation, 
\beq
\frac{\partial \wep_z}{\partial z}
+\frac{\bp}{\ks}\cdot\nabla_\bx \wep_z
+\frac{\ks}{\ep}\cv \wep_z=0
\label{wig}
\eeq
with the initial data
\beq
\label{pure}
W_0(\bx,\bk)=\frac{1}{(2\pi)^d}\int e^{i\bk\cdot\by}
\Psi_0(\bx-\frac{\gamma\by}{2}
)\Psi_0^*(\bx+\frac{\gamma\by}{2})d\by \,,
\eeq
where the operator $\cv$ is formally given as
\beq
\label{L}
\cv\wep_z&=&
i\int e^{i\bq\cdot\bx}
\gamma^{-1}\lt[\wep_z(\bx,\bp+\gamma\bq/2)-\wep_z(\bx,\bp-\gamma\bq/2)\rt]
\widehat{V}(\frac{z}{\ep^2},d\bq)\\
&=&2\gamma^{-1}\int\wepz(\bx,\gamma \bq/2)
\hbox{Im}\lt[e^{-i2\gamma^{-1}\bp\cdot\bx}e^{i\bq\cdot\bx}
\widehat{V}(\frac{z}{\ep^2},d\bq)\rt].
\nn
\eeq

We will use
the following definition of the Fourier transform 
and inversion:
\beqn
\cF f(\bp)&=&\frac{1}{(2\pi)^d}\int e^{-i\bx\cdot\bp}
f(\bx)d\bx\\
\cF^{-1} g(\bx)&=&\int e^{i\bp\cdot\bx} g(\bp) d\bp.
\eeqn
When making a {\em partial}  (inverse) Fourier transform on
a phase-space function we will write $\cF_1$ (resp. $\cF^{-1}_1$) 
and
$\cF_2$ (resp. $\cF^{-1}_2$) 
to denote the  (resp. inverse) transform w.r.t. $\bx$ and $\bp$
respectively.

A useful way of analyzing $\cv \wepz$ as formally given in (\ref{L})  is to look
at its partial inverse Fourier transform $\cF^{-1}_2\cv \wepz (\bx,\by)$ 
acting on 
\beq
\cF^{-1}_2\wepz(\bx,\by)
\equiv\int e^{i\bp\cdot\by}\wepz(\bx,\bp)\,d\bp
=\Psi^\ep(\bx+\gamma\by/2)\Psi^{\ep*}(\bx-\gamma\by/2)
\nn
\eeq
in the following completely local manner
\beq
\label{fourier}
\cF^{-1}_2\cv \wepz(\bx,\by)=
-i\gamma^{-1}\delta_\gamma V^\ep_z(\bx,\by)
\cF^{-1}_2 \wepz(\bx,\by)
\eeq
where
\beq
\label{diff}
\delta_\gamma V^\ep_z(\bx,\by)&\equiv&
V^\ep_z(\bx+\gamma\by/2)-
V^\ep_z(\bx-\gamma\by/2)\\
V^\ep_z(\bx)&=&V_{z/\ep^2}(\bx).
\eeq
Hereby we define for every
realization of $V^\ep_z$
the operator $\cv$ to act on a  phase-space test
function
$\theta$ as
\beq
\cv\theta(\bx,\bp)
\equiv  -i\gamma^{-1}\cF_2\lt[\delta_\gamma
V^\ep_z(\bx,\by) \cF^{-1}_2
\theta(\bx,\by)\rt]
\label{cv}
\eeq
with the difference operator $\delta_\gamma$ given by  (\ref{diff})
for any test function $\theta \in \cS$ where
\[
\cS=\lt\{\theta (\bx,\bp) \in L^2(\IR^{2d}); \pft\theta(\bx,\by)\in C^\infty_c
(\IR^{2d})\rt\}.
\]
We note that $\cv$ is skew-symmetric and
real (i.e. mapping real-valued
functions to real-valued functions).
In this paper we consider the
weak formulation  of the Wigner-Moyal  equation:
To find $W^\ep_z\in D([0, \infty); L^2(\IR^{2d}))$
such that
$\|W^\ep_z\|_2\leq \|W_0\|_2, \forall z>0,$ and 
\beq
\lan W_z^\varepsilon, \theta\ran - \lan W_0,
\theta \ran &=&
\ks^{-1}\int_0^z \lan W_s^\vas, \pdgx \theta\ran ds
+\frac{\ks}{\vas}\int_0^z \lan W_s^\vas, {\cL}^\ep_s
\theta
\ran ds. 
\label{weak}
\eeq

\begin{remark}
Since Eq. (\ref{weak}) is linear , the existence of weak solutions can
be established straightforwardly by  the weak-$\star$ compactness 
argument. Let us briefly comment on this.
First, we introduce truncation $N<\infty$
\[
V_N(z,\bx)=V(z,\bx) ,\quad |V(z,\bx)|< N
\]
and zero otherwise.
Clearly, for such bounded $V_N$ the corresponding operator $\cv$ 
is a bounded self-adjoint operator on $L^2(\IR^{2d})$. Hence the
corresponding Wigner-Moyal equation preserves the $L^2$-norm
of the initial data and produces a sequence of $L^2$-bounded
weak solutions. Passing to the limit $N\to \infty$ we obtain
a $L^2$-weak solution for the original Wigner-Moyal equation
if
$V$ is locally square-integrable as is assumed here.  However,
due to the weak limiting procedure, there is no guarantee that
the $L^2$-norm of the initial data is preserved in the limit.

We will
not address the uniqueness of solution for the Wigner-Moyal
equation (\ref{weak}) but we will show that  as $\ep\to 0$
any sequence of weak solutions to eq. (\ref{weak}) converges
in a suitable sense to the unique solution of a martingale
problem (see Theorem~1 and 2).

\end{remark}

\subsection{Liouville equation}
In the geometric optics limit $\gamma\to 0$, if
one takes the usual WKB-type initial condition
\[
\Psi(0,\bx)=A_0(\bx)e^{i S(\bx)/\gamma}
\]
then the Wigner
 distribution formally tends to the WKB-type  distribution
 \beq
\label{wkb}
W_0(\bx,\bp)=|A_0|^2\delta(\bp-\nabla S(\bx))
 \eeq
which satisfies $\cF_2^{-1} W_0\in L^\infty(\IR^{2d})$.
It has been shown  \cite{Be} that the primitive WKB-type distribution
(\ref{wkb}) can {\em not} arise from the geometrical optics
limit ($\gamma\to 0$) from  any {\em pure} state Wigner
distribution as given by (\ref{pure}) but rather from a {\em mixed}
state Wigner distribution of the form
\beq
\label{mix}
W_0(\bx,\bk)=\frac{1}{(2\pi)^d}\int\int e^{i\bk\cdot\by}
\Psi_0(\bx-\frac{\gamma\by}{2};\alpha
)\Psi_0^*(\bx+\frac{\gamma\by}{2};\alpha)d\by dP(\alpha) \,,
\eeq
where $P(\alpha)$ is a probability distribution of a family
of states $\Psi_0^\alpha$ parametrized by  
$\alpha$. The mixed state Wigner distributions  generally give
rise to
 a smeared initial
condition, i.e. $W_0(\bx,\bp) \in L^2(\IR^{2d})$ even in the geometrical optics
limit. This, instead of the WKB type, is the kind of
initial conditions considered in this paper.

When  acting on the test function space $\cS$,    $\cv$ as given by 
(\ref{cv}) has the following 
limit 
\beq
\label{cv2}
\label{liou}
\lim_{\gamma\to 0}\cv \theta (\bx,\bp)=-\cF_2\lt[\nabla_\bx V_z(\bx)\cdot 
\lt[i\by\cF^{-1}_2
\theta(\bx,\by)\rt]\rt]=-\nabla_\bx V_z(\bx)\cdot\nabla_\bp
\theta(\bx,\bp)
\eeq
in the $L^2$-sense
for all $\theta\in\cS$ and all locally square-integrable $V_z$.
Hence the Wigner-Moyal equation  (\ref{weak}) formally
becomes in the limit $\gamma\to 0$ the Liouville equation in the weak
formulation 
\beq
\label{liou'}
\lan W_z^\varepsilon, \theta\ran - \lan W_0,
\theta \ran &=&
\ks^{-1}\int_0^z \lan W_s^\vas, \pdgx \theta\ran ds
-\frac{\ks}{\vas}\int_0^z \lan W_s^\vas, \nabla_\bx
V_s\cdot\nabla_\bp
\theta
\ran ds,\quad\forall\theta\in \cS.
\eeq

The same weak-$\star$ compactness argument as described in
Remark~1 establishes the existence of $L^2$-weak solution of the
Liouville equation except  now that the operator (\ref{liou}) is
unbounded  and requires local square integrability of $\nabla V_z(\cdot)$.  We
will show that as $\ep\to 0$ any sequence of weak solutions  of the
Wigner-Moyal equation with any
$L^2$-initial condition converge  as $\ep, \gamma \to 0$ in a
suitable sense to the unique solution of a martingale problem
associated with the  Gaussian white-noise model of the Liouville
equation (see Theorem~2).

In addition to the limit $\ep \to 0$ we shall also let
$\rho\to \infty$ and $\eta\to 0$ simultaneously.
We first study  the case
$\rho\to\infty$, but $\eta$ fixed, as
$\ep\to 0$. This means that the Fresnel length
is comparable to the outer scale.
Then we study the narrow beam regime
$\eta\to 0$ where the Fresnel length is 
in the middle of the inertial-convective subrange.

\section{Formulation and main results}
\subsection{Martingale formulation}

The tightness result (see below) implies that
for $L^2$ initial data 
the limiting measure $\IP$ is supported
in $L^2([0,z_0];L^2(\IR^{2d}))$.
For tightness as well as identification of the limit,
 the following infinitesimal operator  $\cA^\ep$ will play an important role.
Let $V^\vas_z\equiv V(z/\ep^2,\cdot)$ and $z_0<\infty$ be any
positive number. Let
$\mathcal{F}_z^\vas$ be the
$\sigma$-algebras generated by $\{V_s^\vas, \, s\leq t\}$  and
$\mathbb{E}_z^\vas$ the corresponding conditional expectation w.r.t. $\cF^\ep_z$.
Let $\cM^\ep$ be the space of measurable function 
adapted to $\{\cF^\ep_z, 
z\in \IR\}$  such that $\sup_{z<z_0}\IE|f_z|<\infty$.
We say $f_z\in \cD(\cA^\ep)$, the domain of $\cA^\ep$, and
$\cA^\ep f_z=g_z$ if $f_z,g_z\in \cM^\ep$ and for
$f^\delta_z\equiv\delta^{-1}[\IE^\ep_z f_{z+\delta}-f_z]$
we have
\beqn
\sup_{z,\delta >0}\IE|f^\delta_z|&<&\infty\\
\lim_{\delta\to 0}\IE|f^\delta_z-g_z|&=&0,\quad\forall t.
\eeqn
Consider a special class of admissible functions
 $f_z=f(\lan W_z^\vas, \theta\ran),
 f'_z=f'(\lan W_z^\vas, \theta\ran),
 \forall f\in C^\infty(\IR)$
we have the following expression
from (\ref{weak}) and the chain rule 
\beq
\label{gen}
 \cA^\vas
f_z
&=&f'_z\lt[ \frac{1}{\ks} \lan W_z^\vas,
{\bp}\cdot \nabla_\bx\theta\ran + \frac{\ks}{\vas} \lan W_z^\vas,
\cv\theta\ran\rt].
\eeq
 A main property of $\cA^\ep$ is
that 
\beq
\label{12.2}
f_z-\int^z_0 \cA^\ep f_s ds\quad\hbox{is a  $\cF^\ep_z$-martingale},
\quad\forall f\in \cD(\cA^\ep).
\eeq
Also,
\beq
\label{mart}
\IE^\ep_s f_z-f_s=
\int^z_s \IE^\ep_s \cA^\ep f_\tau d\tau\quad \forall s<z
\quad\hbox{a.s.}
\eeq
(see \cite{Kur}). Note that the process $W^\ep_z$ is not
Markovian and $\cA^\ep$ is not its generator.
We denote by $\cA$ the infinitesimal operator corresponding
to the unscaled process $V_z(\cdot)=V(z,\cdot)$.

\subsection{The white-noise models}
Now we formulate the solutions for the Gaussian
white-noise model as the solutions to the corresponding martingale problem:
Find the law of $W_z$ on
the subspace of $D([0,\infty);L^2_{w}(\IR^{2d}))$
whose elements have the initial condition 
$W_0(\bx,\bp)\in L^2(\IR^{2d})$
such that
 \beqn
&&f(\lan W_z,\theta\ran)-\int_0^z
 \bigg\{f'( \lan W_s,\theta\ran)\left[\frac{1}{\ks}\lan
 W_s,\pdgx\theta\ran+{\ks^2}
 \lan W_s,\overline{\cQ}_0\theta\ran\right]
  + {\ks^2}f''(\lan W_s,\theta\ran) 
  \lan W_s, \overline{\cK}_{\theta}W_s\ran\bigg\}\,ds\\
\nonumber&&\hbox{{\em is a martingale
for each} $f\in C^\infty(\IR)$}
\eeqn
with
\beq
\label{r10}
 \overline{\cK}_{\theta}W_s
  &=&\int \overline{\cQ}(\theta\otimes
\theta)(\bx,\bp,\by,\bq)W_s(\by,\bq)\,d\by d\bq.
\eeq
Here, in the case of the white-noise model for the  Wigner-Moyal
equation (Theorem~1),  the covariance operators $\overline{\cQ},
\overline{\cQ}_0$ are defined as
\beq
\overline{\cQ}_0\theta&=&
\int \Phi^\infty_\eta(\bq)
 \gamma^{-2}\lt[-2\theta(\bx,\bp)+
  \theta(\bx,\bp-\gamma\bq)
  + \theta(\bx,\bp+\gamma\bq)\rt] {d\bq}.
  \label{102}\\
   \label{101}
 \overline{\cQ}(\theta\otimes \theta)(\bx,\bp,\by,\bq)&=&
 \int e^{i\bq'\cdot(\bx-\by)}
 \Phi^\infty_\eta(\bq')\gamma^{-2}\lt[\theta(\bx,\bp-\gamma\bq'/2)-
  \theta(\bx,\bp+\gamma\bq'/2)\rt]\\
  &&\quad
  \nn
  \hspace{3cm}
 \times \lt[\theta(\by,\bq-\gamma\bq'/2)-
  \theta(\by,\bq+\gamma\bq'/2)\rt]{d\bq'}
\eeq
and, 
in the case of the white-noise model for the  Liouville equation
(Theorem~2),
\beq
\label{101'}
\overline{\cQ}_0\theta(\bx,\bp)&=&
\Delta_\bp \theta (\bx,\bp)\int \Phi^\rho_\eta(\bq)|\bq|^2\,d\bq
     \\
     \label{102'}
 \overline{\cQ}(\theta\otimes \theta)(\bx,\bp,\by,\bq)&=&
  \nabla_\bp\theta(\bx,\bp)\cdot\lt[ \int e^{i\bq'\cdot(\bx-\by)}
   \Phi^\rho_\eta(\bq')\bq'\otimes\bq'
      {d\bq'}\rt]\cdot  \nabla_\bq\theta(\by,\bq),\\
      \nn &&\quad\eta\geq 0,
      \,\,\rho<\infty
     \eeq
with  the spectral density 
   $\Phi^\infty_\eta(\bq)$
    given by
       \[
          \Phi^\infty_\eta(\bq)=
	  \lim_{\rho\to \infty}
	  \Phi^\rho_\eta(\bq)\equiv
	     \lim_{\rho\to \infty}\Phi_{\eta,\rho}(0,\bq),\quad
	     \eta\geq 0.
	        \]
Note that the operators $\overline{\cQ}$
and $\overline{\cQ}_0$
are well-defined
for any test function $\theta \in \cS$ 
in the former case for any $H\in (0,1), \eta>0$
or 
$\eta= 0, 
H\in(0, 1/2)$, and in the latter case for $H\in (0,1),
0\eta<\rho <\infty$ or $H\in (0,1/2), 0=\eta<\rho<\infty$
or
$H\in (1/2,1), 0<\eta<\rho=\infty$.

To see that  (\ref{r10})-(\ref{101})  is square-integrable and
well-defined for any $L^2(\IR^{2d})$-valued process $W_z$,  we apply
$\cF^{-1}_2$ to (\ref{r10}) and obtain
\beq\label{well}
\cF^{-1}_2 \overline{\cK}_{\theta}W_s(\bx, \bu) &=&
\cF^{-1}_2\theta(\bx,\bu) \int
e^{i\bq'\cdot(\bx-\by)}\Phi^\infty_\eta(\bq')
\gamma^{-2} \lt[ e^{i\gamma \bq'\cdot\bu/2}-
e^{-i\gamma\bq'\cdot\bu/2}\rt]\\
&& \hspace{1cm}\times \lt[\theta(\by,\bq-\gamma\bq'/2)-
  \theta(\by,\bq+\gamma\bq'/2)\rt] W_z(\by,\bq)d\by d\bq
{d\bq'}.\nn\\
\nn &=& (2\pi)^{2d}\cF^{-1}_2\theta(\bx,\bu) 
\int \cF^{-1}_2\theta(\by, \by')\lt[
\cF^{-1}_2 W_z(\by,\by')-\cF^{-1}_2 W_z(\by,-\by')\rt]\nn\\
&&\hspace{.3cm}\times\int
e^{-i\by'\cdot\bq'}e^{i\bq'\cdot(\bx-\by)}\Phi^\infty_\eta(\bq')
\gamma^{-2} \lt[ e^{i\gamma \bq'\cdot\bu/2}-
e^{-i\gamma\bq'\cdot\bu/2}\rt]d\bq' d\by d\by'. \nn
\eeq
The integral on the right side of (\ref{well}) is bounded over compact
sets of
$(\bx,\bu)$ because  $\theta\in \cS$, $W_z\in L^2(\IR^{2d})$
and the function 
\[
\Phi^\infty_\eta(\bq')\lt[ e^{i\gamma \bq'\cdot\bu/2}-
e^{-i\gamma\bq'\cdot\bu/2}\rt]
 \]
is integrable in $\bq'\in \IR^{d}$ and the associated integral is bounded
over compact sets of $\bu$ for any $H\in (0,1), \eta>0$
or 
$\eta= 0, 
H<1/2$.  Hence the function on the right side
of (\ref{well}) has a compact
support and  is square-integrable.
Similarly, one can show that (\ref{102})-(\ref{102'})
is well defined for $H\in (0, 1), \rho<\infty$
or $H> 1/2, \rho=\infty$.

In view of the martingale problem  the white-noise model  is an
infinite-dimensional Markov process with the generator given by
\[
\bar{\cA} f_z\equiv f'_s\left[\frac{1}{\ks}\lan
 W_z,\pdgx\theta\ran+{\ks^2}\bar{A}_1( W_z)\right]
 + {\ks^2}f''_z\bar{A}_2( W_z).
\]
This Markov process $W_z$ can also be formulated as
solutions to  the It\^{o}'s equation
 \beq
\label{ito}
 dW_z
 &=&\left(\frac{-1}{\ks}\pdgx +
 {\ks^2}\overline{\cQ}_0\right)
 W_z\,dz+{\ks}d\cbz W_z,\quad W_0(\bx)
 \in L^2(\IR^{2d})
 \eeq
or as the Stratonovich's equation
\beqn
 dW_z
 &=&\frac{-1}{\ks}\pdgx +{\ks}d\cbz \circ W_z,\quad W_0(\bx)
  \in L^2(\IR^{2d})
 \eeqn
  where $\cbz$ is the operator-valued Brownian motion with
  the covariance operator $\overline{\cQ}$, i.e.
  \bean
\IE\lt[d\cbz \theta (\bx,\bp) d\bar{\cB}_{z'}\theta(\by,\bq)\rt]
&=&\delta(z-z')\overline{\cQ}(\theta\otimes\theta)(\bx,\bp,\by,\bq)d z dz'.
\eean
Eq. (\ref{ito}) should be solved  in the space
$D([0,\infty); L^2_w(\IR^{2d}))$, namely, to
find $W_z\in D([0,\infty); L^2_w(\IR^{2d}))$ such that
for all $\theta\in L^2(\IR^{2d})$
\beq
\label{ito2}
d \lan W_z,\theta\ran
 &=&\lan W_z, \left(\frac{1}{\ks}\pdgx +
 {\ks^2}\overline{\cQ}_0\right)
\theta\ran\,dz+ \ks \lan W_z, d\cbz \theta\ran,\quad W_0(\bx)
 \in L^2(\IR^{2d}).
 \eeq
Our results show that
the solution to (\ref{ito2}) exists,  is unique
and 
satisfies the $L^2$-bound
\[
\|W_z\|_2\leq \|W_0\|_2
\]
(cf.  Theorem~1,  2, Remark~1, 3 and  Section~\ref{npt}).

In view of (\ref{101}), (\ref{102}),  (\ref{101'}) and (\ref{102'}) we
can interpret the white-noise limit $\ep\to 0$ as giving
rise to a white-noise-in-$z$ potential $V_z^*$ whose
spectral density is bounded from above by
\[
K^*(\eta^2+|\bk|^2)^{-H^*-d/2}
\]
for some constant $K^*<\infty$ with
the effective H\"older exponent $H_*=H+1/2$
by observing that
\beq
\label{whn-wig}
\lim_{\ep\to 0}\cv\theta(\bx,\bp) &=&-i \cF_2\lt[ 
\gamma^{-1} \delta_\gamma V^*_z(\bx,\by)
\cF^{-1}_2\theta(\bx,\by)\rt],\quad \forall \theta\in \cS\\
\label{liou2}
\lim_{\ep, \gamma\to 0}\cv\theta(\bx,\bp) &=&\nabla_\bx
V^*_z(\bx)
\cdot\nabla_\bp\theta(\bx,\bp),\quad \forall \theta\in \cS
\eeq
in the  mean square sense. 

\subsection{White-noise models with large-scale inhomogeneities}
\label{inhom2}
First we consider the case of deterministic, large-scale inhomogeneities of  a
multiplicative type which has $\mu$,  given by (\ref{inhom}), as a
bounded smooth function
$
 \mu=\mu(z,\bx).
$
The resulting limiting process can be described  analogously as above
except with the term $\Phi^\infty_\eta$ replaced
by
\beqn
\Phi^\infty_\eta(\bk)&\longrightarrow&\mu(z,\bx)\mu(z,\by)\Phi^\infty_\eta(\bk),\quad
\hbox{in}\quad \overline{\cQ}\\
\Phi^\infty_\eta(\bk)&\longrightarrow&\mu^2(z,\bx)\Phi^\infty_\eta(\bk),\quad
\hbox{in}\quad \overline{\cQ}_0.
\eeqn
As a consequence the operator $\overline{\cQ}_0 $
is no longer of convolution type.

\commentout{
More generally we can consider the following slowly modulated
refractive index field in its spectral representation
\[
V_z^\ep(\bx)=\int e^{i\bk\cdot\bx} e^{i\xi z\ep^{-2}}
\hat{V}(d\xi, d\bk; \xvec), \quad \xvec=(z,\bx)
\]
where the modulated spectral density $\hat{V}(d\xi, d\bk;
\xvec)$ satisfies 
\[
\IE\lt[ \hat{V}(d\xi, d\bk, \xvec) 
\hat{V}(d\xi', d\bk', \xvec')\rt]=
\Phi_{(\eta,\rho)}(\xi, \bk; \xvec, \xvec') d\xi d\xi' d\bk d\bk'.
\] In this case the resulting kernel in the
definition of
$\overline{\cQ} $ and $\overline{\cQ}_0$ should change in the
following way 
\beqn
\Phi^\infty_\eta(\bk)&\longrightarrow&\Phi^\infty_\eta(\bk;
z,\bx, z,\by)=\Phi_{(\eta,\infty)}(0,\bk;z,\bx, z,\by)\quad
\hbox{in}\quad
\overline{\cQ}\\
\Phi^\infty_\eta(\bk)&\longrightarrow&\Phi^\infty_\eta(\bk;
z,\bx,
z,\bx)=\phi_{(\eta,\infty)}(0,\bk;z,\bx,z,\bx),\quad\hbox{in}\quad
\overline{\cQ}_0.\\
\eeqn
}

Next we add a slowly varying  smooth deterministic background
$V_0(z,\bx)$ to the rapidly fluctuating field
$\ep^{-1}\mu(z,\bx) V(\ep^{-2}z,\bx)$. Namely we have
\[
V_0(z,\bx)+\frac{\mu(z,\bx)}{\ep} V(\frac{z}{\ep^2},\bx)
\]
as the potential term in the parabolic wave equation (\ref{para}).

The resulting martingale problem has an additional term
 \beq
 \label{inhom3}
-\int_0^z
 \ks \lan W_s, \cL_0 \theta\ran
  \,ds
\eeq
in  the martingale formulation
where $\cL_0\theta$ has the
form
\beq
\nn
\cL_0\theta(\bx,\bp)&=&
i\int e^{i\bq\cdot\bx}
\gamma^{-1}\lt[\theta(\bx,\bp+\gamma\bq/2)-\theta(\bx,\bp-\gamma\bq/2)\rt]
\widehat{V}_0({z},d\bq)\\
&\equiv&
  -i\gamma^{-1}\cF_2\lt[
(V_0(\bx+\gamma\by/2)-
V_0(\bx-\gamma\by/2)) \cF^{-1}_2
\theta(\bx,\by)\rt]\label{inhom4}
\eeq
for $\gamma>0$ fixed in the limit, and
the form
\beq
\label{inhom4'}
\cL_0\theta(\bx,\bp)=-\nabla_\bx V_0(z, \bx)\cdot\nabla_\bp
\theta(\bx,\bp)
\eeq
in the case of $\gamma\to 0$.

\subsection{Multiple-point correlation functions of the limiting model}
\label{npt}
The martingale solutions of the limiting models  are uniquely
determined by their $n$-point correlation functions
which satisfy a closed set of evolution equations.

Using the  function $f(r)=r^n$ in the martingale formulation
and taking expectation,
we arrive after some algebra the following equation
\beq
\label{npt'}
\frac{\partial F^{(n)}}{\partial z}&=&
\frac{1}{\ks}\sum_{j=1}^{n}\bp_j\cdot\nabla_{\bx_j} F^{(n)}
+{\ks^2}\sum_{j=1}^{n}\overline{\cQ}_0(\bx_j,\bp_j)F^{(n)}
+{\ks^2}\sum_{{j,k=1}\atop {j\neq k}}^{n}\overline{\cQ}
(\bx_j,\bp_j,\bx_k,\bp_k)
F^{(n)}
\eeq
for the $n-$point correlation function
\[
F^{(n)}(z,\bx_1,\bp_1,\dots,\bx_n,\bp_n)\equiv
\IE\lt[ W_z(\bx_1,\bp_1)\cdots W_z(\bx_n,\bp_n)\rt]
\]
where $\overline{\cQ}_0(\bx_j,\bp_j)$ is the operator
$\overline{\cQ}_0$ acting on the variables $(\bx_j,\bp_j)$
and $\overline{\cQ}(\bx_j,\bp_j,\bx_k,\bp_k)$ is the operator
$\overline{\cQ}$ acting on the variables $(\bx_j,\bp_j,\bx_k,\bp_k)$,
namely
\beqn
\lefteqn{{\cQ}(\bx_j,\bp_j,\bx_k,\bp_k)F^{(n)}(\bx_i,\bp_i) }\\
&=&
\IE\lt\{ \lt[\prod_{i\neq j,k}W(\bx_i,\bp_i)\rt] \int e^{i
\bq'\cdot(\bx-\by)} 
     \Phi_{(\eta,\infty)}(0,\bp) \gamma^{-2} \right.\\
	& & \left.\times
 [W(\bx_j, \bp_j-\gamma \bq/2) - W(\bx_j, \bp_j+\gamma
\bq/2)]
 [W(\bx_k, \bp_k-\gamma \bq/2) - W(\bx_k,
\bp_k+\gamma
\bq/2)]
	 \,\,d\bq \rt\}. 
\eeqn

Eq. (\ref{npt'}) can be more conveniently written as
\beq
\label{npt2}
\frac{\partial F^{(n)}}{\partial z}&=&
\frac{1}{\ks}\sum_{j=1}^{n}\bp_j\cdot\nabla_{\bx_j} F^{(n)}
+{\ks^2}\sum_{{j,k=1}}^{n}\overline{\cQ}
(\bx_j,\bp_j,\bx_k,\bp_k)
F^{(n)}
\eeq
with the identification $\overline{\cQ}
(\bx_j,\bp_j,\bx_j,\bp_j)=\overline{\cQ}_0(\bx_j,\bp_j)$.
The operator  
\beq
\label{35'}
\sum_{{j,k=1}}^{n}\overline{\cQ}
(\bx_j,\bp_j,\bx_k,\bp_k)
\eeq
 is a non-positive symmetric operator. 
We note that the mean Wigner distribution can be exactly
solved for from Eq. (\ref{npt2}) for $n=1$ \cite{tire} and has a
number of interesting applications in optics including  time
reversal.  The 2-nd moment equation $n=2$ is related to the
problem of scintillation \cite{St} (see, e.g., \cite{BO}).

The uniqueness for 
eq. (\ref{npt'}) with any initial data
\[
F^{(n)}(z=0,\bx_1,\bp_1,\dots,\bx_n,\bp_n)=
\IE\lt[ W_0(\bx_1,\bp_1)\cdots W_0(\bx_n,\bp_n)\rt],\quad W_0\in L^2(\IR^{2d})
\]
in the case of the Wigner-Moyal equation
can be easily established by observing that
the operator given by (\ref{35'}) is self-adjoint.
In the case of the Liouville equation, eq. (\ref{npt2}) can be
more explicitly written as the advection-diffusion equation
on the phase space
\beq
\label{npt22}
\frac{\partial F^{(n)}}{\partial z}&=&
\frac{1}{\ks}\sum_{j=1}^{n}\bp_j\cdot\nabla_{\bx_j} F^{(n)}
+\frac{\ks^2}{4}\sum_{{j,k=1}}^{n}{D}(\bx_j-\bx_k):\nabla_{\bp_j}
\nabla_{\bp_k}
F^{(n)}
\eeq
with
\beqn
{\bD}(\bx_j-\bx_k)&=&
\int e^{i\bq'\cdot(\bx_j-\bx_k)}
   \Phi^\rho_\eta(\bq')\bq'\otimes\bq'
         {d\bq'}\\
{D}(0)&=&\int
   \Phi^\rho_\eta(\bq')|\bq'|^2
            {d\bq'}
\eeqn
with $\eta\geq 0$ where
$D(0)$ is the Stratonovich correction term.
In the worst case scenario the diffusion coefficient ${D}(0)$ diverges
as $\rho\to \infty$ but always well-defined as $\eta\to 0$ for
$H<1/2.$ Moreover
the diffusion operator 
\[
\sum_{{j,k=1}}^{n}{\bD}(\bx_j-\bx_k):\nabla_{\bp_j}
\nabla_{\bp_k}
\]
is an  essentially self-adjoint positive operator on $C_c^\infty(\IR^{2nd})$
due to the sub-Lipschitz growth of the square-root
 of ${\bD}(\bx_k- \bx_k)$ at large $|\bx_j|, |\bx_k|$
\cite{Da}.

\subsection{Assumptions and properties of the refractive index field}
\label{assumption}
As mentioned in the introduction,
we assume that $V_z(\bx) $ is a square-integrable, $z$-stationary,
$\bx$-homogeneous process  with a spectral density satisfying
the upper bound (\ref{power}).  

Let $\rho(t)$ be a non-negative (random or deterministic)
function such that
\beq
\lt|\IE \lt[\IE_z[V_s(\bx)] \IE_z[V_t(\by)]\rt]\rt|
&=&\lt| \IE \lt[\IE_z[V_s(\bx)] V_t(\by)\rt]\rt|\nn\\
&\leq& r(s-z)r(t-z) \IE\lt[V_t^2\rt],\quad
\forall s, t\geq z,\forall
\bx,\by \in \IR^d.
\label{decay}
\eeq

An obvious candidate for $r(t)$ is the
 correlation coefficient defined as follows.
Let $\cF_z $ and $\cF^+_z$  be the sigma-algebras generated by
$\{V_s:  \forall s\leq z\}$ and $\{V_s: \forall s\geq  z\}$,
respectively. 
The correlation coefficient $\corr(t)$ is given by
\beq
\label{correl}
\corr(t)=\sup_{h\in \cF_z\atop \IE[h]=0,
\IE[h^2]=1}
\sup_{
g\in \cF_{z+t}^+\atop \IE[g]=0,
\IE[g^2]=1}\IE\lt[h g\rt].
\eeq
\begin{lemma}
\label{newlemma1}
The  correlation coefficient $\corr(t)$ as given by
(\ref{correl}) satisfies the inequality (\ref{decay}). 
\end{lemma}
\begin{proof}
Let
\[
h_s(\bx)=\IE_z[V_s(\bx)],\quad g_t(\bx)= V_t(\bx).
\]
Clearly 
\beqn
h_s
&\in & L^2(P,\Omega, \cF_{z})\\
g_t &\in& \in L^2(P,\Omega,
\cF^+_{t})
\eeqn
and their second moments are uniformly bounded
in $\bx$ since
\beqn
\IE[h_s^2](\bx)&\leq& \IE[g_s^2](\bx)\\
\IE[g_s^2](\bx) &=& \int
\Phi(\xi,\bq) d\xi d\bq
\eeqn

 From the definition
(\ref{correl}) we have
\beqn
\lt|\IE[h_s(\bx) h_t(\by)]\rt|&=
\lt|\IE\lt[h_s g_t\rt]\rt|
&\leq \corr(t-z) \IE^{1/2}\lt[h_s^2(\bx)\rt]
\IE^{1/2}\lt[g_t^2\rt].
\eeqn
Hence by  setting $s=t$ first and the
Cauchy-Schwartz inequality we have
\beqn
\IE\lt[h_s^2\rt(\bx)]&\leq&\corr^2(s-z)\IE[g_t^2]\\
\IE\lt[ h_s(\bx) h_t(\by)\rt]
&\leq& \corr(t-z) \corr(s-z)
\IE[g_t^2],\quad \forall s,
t\geq z,\forall \bx, \by.
\eeqn
\end{proof}

We assume
\begin{assumption}
\label{ass1}
\begin{itemize}\item[(i)]
The function 
$r(t)$ in (\ref{decay}) satisfies 
\[
\int_0^\infty\int^\infty_0 \IE[r(s)r(t)] ds dt
<\infty.
\]
\item[(ii)] 

\end{itemize}
\end{assumption}

\commentout{**********
\begin{corollary}
\label{cor1}
The formula 
\beq
\label{new}
\tilde{ V}_z(\bx)=\lim_{M\to \infty}
\int^{z+M}_z\IE_z\lt[
V_s(\bx)\rt]\,\,ds -\int^M_0\IE_0\lt[ V_s(\bx)\rt]\,\,ds
\eeq
defines 
a square-integrable  $\bx$-homogeneous  process with
$\tilde{V}_0(\bx)=0$. 
\end{corollary}
*************}
\begin{corollary}
\label{cor1}
Assumption~1 implies that the formula 
\beq
\label{new}
\tilde{ V}_z(\bx)=
\int^\infty_0\IE_0\lt[ V_s(\bx)\rt]\,\,ds
\eeq
defines 
a square-integrable  $\bx$-homogeneous  process. 
\end{corollary}
\begin{proof}
We only need to verify its second moment is finite:
\beqn
\IE\lt[ \tilde{V}^2_z(\bx)\rt]
&=&\IE\lt[\int^\infty_z\int^\infty_z
\IE_z[V_s(\bx)]\IE_z[V_t(\bx)] ds dt\rt]\\
&=&\IE\lt[\int^\infty_z\int^\infty_z 
\IE_z[V_s(\bx)]V_t(\bx) ds dt\rt]\\
&\leq&\IE\lt[\int^\infty_0\int^\infty_0 
r(s)r(t)ds dt\rt] \IE[V_z^2]
\eeqn
which is finite by Assumption~1.
\end{proof}
One can adopt other alternative mixing
coefficients to get the above results and
higher order moment estimates, see Appendix
A. Hereafter we will mainly focus on
the correlation coefficient as it
is most convenient to work with
in the Gaussian case and we shall
write explicitly the dependence of
the correlation coefficient on  $\eta, \rho$
as $r_{\eta,\rho}(t)$.

In the Gaussian case
the correlation coefficient
$\corr(t)$ equals the linear correlation
coefficient  given by
\beq
\label{corr}
\corr(t) &=&\sup_{g_1, g_2}
 \int R(t-\tau_1-\tau_2,\bk) g_1(\tau_1,\bk)
g_2(\tau_2,\bk)d\bk d\tau_1 d\tau_2
\eeq
where
\[
R(t,\bk)=\int e^{it\xi} \Phi_{(\eta,\rho)}(\xi, \bk) d\xi
\]
and the
supremum is taken over all $g_1, g_2\in L^2(\IR^{d+1})$ which are
supported on $(-\infty, 0]\times
\IR^d$ and  satisfy the constraint
\beq
\label{b.1}
\int R(t-t',\bk) g_1(t,\bk)\bar{g}_1(t',\bk) dt dt' d\bk=
\int R(t-t',\bk) g_2(t,\bk)\bar{g}_2(t',\bk) dt dt' d\bk=1.
\eeq
Alternatively, by the Paley-Wiener theorem we can write
\beq
\label{corr2}
\corr(t) &=&\sup_{f_1, f_2}
 \int e^{i\xi t}f_1(\xi,\bk)f_2(\xi,\bk)
\Phi(\xi,\bk)d\xi d\bk
\eeq
where $f_1, f_2$ are elements of the  Hardy space $\cH^2$
of $L^2(\IR^d; \Phi_{(\eta,\rho)})$-valued
analytic functions in
the upper half
$\xi$-space satisfying the normalization condition
\[
\int |f_j(\xi,\bk)|^2\myphi(\xi,\bk)d\xi d\bk=1, \quad j=1,2.
\]
There are various criteria for the decay rate of the linear correlation coefficients, see \cite{IR}.
 
\begin{corollary}
\label{cor2}
If $V_z$ is a Gaussian random field and its linear correlation
coefficient $\corr(t)$ is integrable,
then $\tilde{V}_z$ is also Gaussian and hence possesses finite
moments of all orders.
\end{corollary}
This follows from the fact that the mapping from
$V_z$ to $\tilde{V}_z$ is a bounded linear operator on
the Gaussian space.

The main property of $\tilde{V}_z$  as a random function is that
\beq
\label{eq-a}
\cA\tilde{ V}_z=- V_z,\quad\hbox{a.s. } \quad z\in \IR.
\eeq
Since $\cA$ commutes with the  shift
in $\bx$ so the appearance of $\bx$ in  eq. (\ref{eq-a}) is
 suppressed.

We  have the following simple relation
\commentout{
\beq
\label{19'}
\IE\lt[\tilde{ V}_z(\bx) V_z(\by)\rt]
&=&\int e^{i(\bx-\by)\cdot \bp}
\int^\infty_0 e^{i\xi s} ds \Phi_{\eta,{\rho}}(\xi,\bp)\,d\xi\,d\bp\\
&=&\int  e^{i(\bx-\by)\cdot \bp}  \lim_{L\to\infty}\int \frac{1}{i\xi}
\lt(e^{iL\xi}-1\rt)
\Phi_{(\eta,\rho)}(\xi,\bp) d\xi d\bp\nn\\
&=&\pi\int  e^{i(\bx-\by)\cdot \bp}  \Phi_{(\eta,\rho)}(0,\bp)
d\bp\nn
\eeq
and similarly
}
\beq
\label{19}
\lim_{\lamb\to\infty} \IE\lt[\tilde{ V}_{z\lamb} (\bx)
V_{z\lamb}(\by)\rt] &=& \lim_{\lamb\to\infty} \int  e^{i(\bx-\by)\cdot
\bp} 
\int \frac{1}{i\xi}
\lt(e^{iz\lamb\xi}-1\rt)
\Phi_{(\eta,\rho)}(\xi,\bp) d\xi d\bp\\
&=&\pi\int  e^{i(\bx-\by)\cdot \bp}  \Phi_{(\eta,\rho)}(0,\bp)
d\bp,\quad\forall z.\nn
\eeq
Define the covariance functions
\beqn
\tilde{B}_z(\bx-\by)&\equiv&\IE\lt[\tilde{ V}_z(\bx)
\tilde{V}_{z}(\by)\rt]
\eeqn
and write
\beqn
\tilde{B}_z(\bx)&=&\int e^{i\bk\cdot\bx}
\tilde{\Phi}_z(\bk)d\bk
\eeqn
 where $\tilde{\Phi}_z(\bk)$ is its spectral density function. 

By the 
properties of the
orthogonal projection  $\IE_z[\cdot]$, we know that
\beq
\label{2.48}
\IE\lt[\IE_z[\hat{V}(A)]\IE_z[\hat{V}(A)]\rt]
\leq \IE\lt[\hat{V}(A)\hat{V}(A)\rt]
=\int_A \Phi_{(\eta,\rho)}(\xi,\bk)d\xi d\bk
\eeq
for every Borel set $A\subset \IR^{d+1}$.

\begin{assumption}
\label{ass3}
(i) The limit $\rinf=\lim_{\rho\to \infty}\corr(t)$
exists and satisfies
\[
\int^\infty_0 \rinf(t) dt= \lim_{\rho\to
\infty}\int^\infty_0 \corr(t)d t<\infty.
\]

(ii) As $\eta\to 0$, there exists a constant $C$ such that
\[
\int^\infty_0 \corr(t) dt\leq C\eta^{-1} \int^\infty_0
r_{1,\rho}(t)dt
\]
\end{assumption}
For the Gaussian case with the von K\'arm\'an spectrum 
(\ref{vk}), a 
straightforward scaling argument shows that
\[
\corr(t)=r_{1,\rho}(\eta t)
\]
which is indepedent of $\rho$. This
 motivates Assumption 2.

Set
\[
\tilde{\Phi}^\ep_z(\bk)\equiv
\tilde{\Phi}_{\ep^{-2}z}(\xi,\bk)
\]
which is the spectral density of  $\tilde{V}_z^\ep(\bx)\equiv
\tilde{V}_{z/\ep^2}(\bx)$.

Define analogously to (\ref{cv})
\beq
\tilde{\cL}^\ep_z\theta(\bx,\bp)
\equiv  -i\gamma^{-1}\cF_2\lt[\delta_\gamma
\tilde{V}^\ep_z(\bx,\by) \cF^{-1}_2
\theta(\bx,\by)\rt]
\label{cvtil}
\eeq
with
\[
\delta_\gamma \tilde{V}^\ep_z(\bx,\by)\equiv
\tilde{V}^\ep_z(\bx+\gamma\by/2)-\tilde{V}^\ep_z
(\bx-\gamma\by/2).
\]

\begin{lemma}
\label{newlemma2}(Appendix B) For each $z_0<\infty$ there exists
a positive constant
$\tilde{C}<\infty$ such that
\beqn
\sup_{|z|\leq z_0\atop |\by|\leq L} \IE\lt[\lt(\delta_\gamma
{V}^\ep_{ z}\rt)^2\rt] (\by) &\leq &\tilde{C}
\gamma^2\lt|\min{(\gamma^{-1},\rho)}\rt|^{2-2H}\\
\sup_{|z|\leq z_0} \IE\lt[\tilde{V}^\ep_{z} (\bx)\rt]^2
&\leq &\tilde{C}
\eta^{-2-2H}
\\
\sup_{|z|\leq z_0 \atop |\by|\leq L}
\IE\lt[\lt(\delta_\gamma\tilde{V}^\ep_{ z}\rt)^2\rt]
(\by) &\leq &\tilde{C}\eta^{-2}\gamma^2 \minrg^{2-2H}\\
\sup_{|z|\leq z_0\atop |\by|\leq L}\lt|\nabla_\by
\IE\lt[\delta_\gamma
\tilde{V}^\ep_z\rt]^2(\by)\rt|&\leq&\tilde{C}\eta^{-2}
\gamma^2\rho^{1-H}\minrg^{1-H}\\
  \sup_{|z|\leq
z_0}\IE\|\pdgx(\cvtil\theta)\|_2^2&\leq&\tilde{C}\eta^{-2}
\rho^{4-2H},\quad \theta\in\cS
  \eeqn
  for all $ H\in (0,1), 
 \ep,\gamma,\eta\leq 1\leq \rho, \bx,\by\in\IR^d$
  where the constant $\tilde{C}$ depends only on $z_0$, $L$
  and $\theta$.
  \end{lemma}

We also need to know the first few moments the random fields
involved.  The case of Gaussian fields motivates the following
assumption of the 
6-th order quasi-Gaussian property. 
\begin{assumption}
\label{ass4}
\beq
\label{ga1}
\sup_{|\by|\leq L}\IE\lt[\delta_\gamma V^\ep_z(\by) \rt]^4&\leq&
C_1  \sup_{|\by|\leq L}\IE^2\lt[\delta_\gamma
V^\ep_z\rt]^2(\by)\\
\label{ga2}
\sup_{|\by|\leq L}\IE\lt[\delta_\gamma \tilde{V}^\ep_z
\rt]^4(\by)&\leq& C_2\sup_{|\by|\leq L} \IE^2\lt[\delta_\gamma
\tilde{V}^\ep_z\rt]^2(\by)\\
\label{ga3}
\sup_{|\by|\leq L}\IE\lt[\lt[\delta_\gamma
{V}^\ep_z \rt]^2\lt[\delta_\gamma
\tilde{V}^\ep_z \rt]^4\rt](\by)&\leq& C_3\lt(\sup_{|\by|\leq L}
\IE\lt[\delta_\gamma V^\ep_z\rt]^2(\by)\rt)\lt(
\sup_{|\by|\leq L} \IE^2\lt[\delta_\gamma
\tilde{V}^\ep_z\rt]^2(\by)\rt)
\eeq
for all $L<\infty$ where the constants $C_1, C_2$ and $C_3$ are independent
of $\ep, \eta, \rho, \gamma$. 
\end{assumption}

From (\ref{cv}) and (\ref{cvtil}) we can form the iteration
of operators $\cv\cvtil$
\[
\cv\cvtil\theta(\bx,\bp)
= -\gamma^{-2}\cF_2\lt[\delta_\gamma V^\ep_z(\bx,\by)
\delta_\gamma
\tilde{V}^\ep_z(\bx,\by)\cF^{-1}_2
\theta(\bx,\by)\rt]
\]
The operator $\cv\cvtil\theta$ is well-defined if  
$\delta_\gamma
V^\ep_z$ and $\delta_\gamma\tilde{V}^\ep_z$  are locally 
square-integrable. Other iterations of $\cv$ and $\cvtil
$ allowed by Assumption 2 can be similarly constructed.

The following estimates can be obtained
from Lemma~2 and Assumption~3.

\begin{corollary} 
\label{cor3}
\beq
\IE\lt[\|\cv\theta(\bx,\bp)\cvtil\theta(\by,\bq)\|_2^2\rt]
\nn
&=&O\lt(\gamma^{-4}\sup_{|\by|\leq L}\IE\lt|\delta_\gamma
V^\ep_z\rt|^2(\by)\IE\lt|\delta_\gamma
\tilde{V}^\ep_z\rt|^2(\by)\rt)
=O\lt(\eta^{-2}\minrg^{4-4H}\rt)\label{gauss1}\\
\IE\lt[\|\cv\cvtil\theta\|_2^2\rt]
\label{gauss2} 
&=&O\lt(\gamma^{-4}\sup_{|\by|\leq L}\IE\lt|\delta_\gamma
V^\ep_z\rt|^2(\by)\IE\lt|\delta_\gamma
\tilde{V}^\ep_z\rt|^2(\by)\rt)=
O\lt(\eta^{-2}\minrg^{4-4H}\rt)\nn\\
\IE\lt[\|\cvtil\cvtil\theta\|_2^2\rt]
\label{gauss3}
&=&O\lt(\gamma^{-4}\sup_{|\by|\leq L}\IE^2\lt|\delta_\gamma
\tilde{V}^\ep_z\rt|^2(\by)\rt)=O\lt(\eta^{-4}
\minrg^{4-4H}\rt)\nn\\
\label{gauss4}
\IE \lt\|\cv\cvtil\cvtil\theta\rt\|^2_2
&=&O\lt(\gamma^{-6}\sup_{|\by|\leq L}\IE^2\lt|\delta_\gamma
\tilde{V}^\ep_z\rt|^2\IE\lt|\delta_\gamma
V^\ep_z\rt|^2\rt)=O\lt(
\eta^{-4}\minrg^{6-6H}\rt)\nn
\eeq
where the constants are independent of $\rho,\eta,
\gamma$ and $L$ is the radius of the ball containing the support
of $\cF^{-1}_2\theta$.  

\end{corollary}

\begin{assumption}
\label{ass5}
For every $\theta\in \cS$, there exists a random
constant
$C_5$ 
such that
\beq
\label{21}
\sup_{z<z_0}\|\delta_\gamma
\tilde{V}^\ep_z\cF^{-1}_2\theta\|_4
 &\leq& \frac{C_5}{\sqrt{\ep}}
\sup_{z\in [0, z_0]\atop
|\bx|,|\by|\leq L}\IE^{1/2}|
\delta_\gamma
\tilde{V}^\ep_z(\bx,\by)|^2,
\quad\forall\theta\in \cS, 
\ep,\eta,\gamma\leq 1\leq
\rho
 \eeq
with $C_5$ possessing finite moments
  and depending
only on $\theta, z_0$, where $L$ is the
radius of the ball containing the support
of $\invf \theta$.
cf. Lemma~2 and
(\ref{g1}). 
\end{assumption}
For a Gaussian random field,
 Assumption~3  is readily
satisfied for any $\alpha\in (0,1)$ by  Lemma~2 and a simple
application of  Borell's  inequality \cite{Ad}
  \beq
\nn
  \sup_{z<z_0}\|\delta_\gamma
\tilde{V^\ep_z}\cF^{-1}_2\theta\|_4
&\leq &\|\cF^{-1}_2\theta\|_4 \sup_{z\in [0,
z_0]\atop |\bx|, |\by|\leq L} |\delta_\gamma 
\tilde{V}^\ep_z(\bx,\by)|\\
   &\leq& C_5
\log{\lt(\frac{z_0}{\vas^2}\rt)}\sup_{z\in
[0,z_0]\atop |\bx|,|\by|\leq L}\IE^{1/2}|\delta_\gamma
\tilde{V^\ep_z}(\bx,\by)|^2
, \quad  \forall \eta,\gamma\leq 1\leq
\rho
\label{g1}
\label{59}
       \eeq
where the random constants  $C_5$
  has a Gaussian-like tail. 

 Note that with $\gamma$ or $\rho$ held fixed 
the first term on the right side of (\ref{21}) is always $O(1)$. 
Compared
to the  corresponding condition  (\ref{g1}) for the Gaussian field
condition (\ref{21}) allows for  certain degree of  intermittency in
the refractive index field.

\commentout{
One can obtain 
similar conditions for the spatial derivatives
of the above quantities, for instance,
\beq
\sup_{z<z_0}\|\nabla_\by\cdot\nabla_\bx\delta_\gamma V^\ep_z\cF^{-1}_2\theta\|^2_2
&\leq &\frac{C_1}{\ep}\rho^{2(1-H)},\quad\mbox{independent of $\eta$}\\
\sup_{z<z_0}\|\nabla_\by\cdot\nabla_\bx
\delta_\gamma \tilde{V^\ep_z}\cF^{-1}_2\theta\|^2_2
&\leq& \frac{C_2}{\ep}, \quad \mbox{independent of $\rho,\eta$.} 
\eeq
}
    
\commentout{
When $\gamma$ is also allowed to vanish  with $\ep$ (Theorem~2 )
we assume analogous 
conditions with stronger dependence on $\rho$: 
for some $\alpha\in (0,1/4)$
\beq
\label{21'}
\limsup_{\gamma\to 0} \sup_{z<z_0} \gamma^{-1}\|\delta_\gamma
\tilde{V}^\ep_z\cF^{-1}_2\theta\|_4
 &\leq&
\frac{C_4}{\sqrt{\ep}}\ep^{-2\alpha}\lt[\eta^{\alpha-H}+\rho^{\alpha-H
}
 \rt]
     \eeq
     where random constants $C_4$ has finite moments.
}

Finally,  we assume that for all $\rho<\infty$
the refractive index field is smooth in the transverse coordinates.

\commentout{*************
\subsection{Example: Gaussian random fields}
\label{example}
By the Karhunen theorem \cite{Ka}
and  the
existence of an integrable spectral density, the random field admits
$V_z$ a moving average representation
\beq
\label{move}
V_z(\bx)=\int \Psi(z-s, \bk) W(d s, d\bk)
\eeq
where $\Psi\in L^2(\IR^{d+1})$, $W(\cdot, \cdot)$ is a
complex orthogonal random measure on $\IR^{d+1}$ such that
\[
\IE|W(\triangle)|^2=|\triangle|
\]
for all Borel sets $\triangle \subset \IR^{d+1}$. 
With
\beqn
\hat{\Psi}(\xi,\bk)&=&\frac{1}{2\pi}\int e^{-i\xi s} \Psi (s,\bk)\nn
\eeqn
 we have the following
relation between the spectral measures $\hat{V}(d\xi,d\bk)$
and $\hat{W}(d\xi,d\bk)$, on one hand, 
\beqn
\label{b}
\hat{V}(d\xi,d\bk)&=& \hat{\Psi}(\xi,\bk)\hat{W}(d\xi,d\bk)
\eeqn
 and, on other hand,
 between the spectral density $\Phi_{(\eta,\rho)}$
 and  the Fourier-transform $\hat{\Psi}$
\beqn
\Phi_{(\eta,\rho)}(\xi,\bk)&=&|\hat{\Psi}(\xi,\bk)|^2.
\eeqn

As a result of Corollary~\ref{cor2}
 $\tilde{V}_z$ is Gaussian
and hence Assumption~2 is satisfied. 
*************}

\subsection{Main theorems}
\begin{theorem}
\label{thm1}
Let $V^\ep_z$ be a $z$-stationary,
$\bx$-homogeneous,  almost surely  smooth, locally bounded random
process with the spectral density satisfying the bound (\ref{power})
and Assumptions 1,2,3. Let $\gamma>0$ be fixed.
\begin{itemize}
\item[(i)]
Let $\eta$ be fixed and  ${\rho}$ be fixed or tend
to $
\infty$ as
$\ep\to 0$ such that
\beq
\label{ass2}
\lim_{\ep\to
0}\ep\rho^{2-H}=0.
\eeq
Then
the weak solution $ W^\ep$ of the Wigner-Moyal equation with
the initial condition $W_0\in L^2(\IR^{2d})$ converges in
law
in the space $D([0,\infty);L^2_w(\IR^{2d}))$  
of $L^2$-valued right continuous processes with left limits
endowed with the Skorohod topology
to that of
the corresponding
Gaussian white-noise model with the covariance operators $\overline{\cQ}$
and $\overline{\cQ}_0$ as given by
(\ref{101}) and (\ref{102}), respectively (see also
(\ref{inhom3}) and (\ref{inhom4})). 
The statement holds true for any $H\in (0,1)$.

\item[(ii)] 
Suppose additionally  that $H<1/2$
and
$\eta=\eta(\ep)\to 0$ such that
\beq
\label{ass}
\lim_{\ep\to
0}\ep\eta^{-1}(\eta^{-1}+\rho^{2-H})=0.
\eeq
Then the same
convergence holds.
\end{itemize}
\end{theorem}
Here and below $L^2_w(\IR^{2d})$ is the space of square integrable
functions on the phase space $\IR^{2d}$ endowed with
the weak topology.

The next theorem concerns a similar convergence
to  the solution of  a Gaussian white-noise model for the Liouville
equation. 
\begin{theorem}

Let $V^\ep_z$ be a $z$-stationary,
$\bx$-homogeneous,  almost surely  smooth, locally bounded random
process with the spectral density satisfying the bound (\ref{power}) 
and Assumptions 1,2,3.

Let $\gamma=\gamma(\ep)\to 0$ as $\ep\to 0$.
Then under  any of the following three sets of conditions
\begin{itemize}
\label{thm2}
\item[(i)]
 ${\rho}<\infty$ and $\eta>0$ held fixed;

\item[(ii)]
 $H>1/2$, $\eta>0$ fixed and $\rho=\rho(\ep)\to \infty$ as  $\ep\to
0$ such that
\beq
\label{rho}
\lim_{\ep\to 0}\ep\rho^{2-H}=0;
\eeq

\item[(iii)]
 $H<1/2$, $\rho<\infty$ fixed
and
$\eta=\eta(\ep)\to 0$ such that
\beq
\label{ass'}
\lim_{\ep\to 0}\ep\eta^{-2}=0;
\eeq
\end{itemize}
the weak solutions $W^\ep$ of the Wigner-Moyal equation
(\ref{wig}) with the initial condition $W_0\in
L^2(\IR^{2d})$ 
converges in
distribution
in the space $D([0,\infty); L^2_w(\IR^{2d}))$ to
the martingale solution  of
the Liouville equation of the
Gaussian white-noise model with the covariance operators $\overline{\cQ}$
and $\overline{\cQ}_0$ as given by (\ref{101'}) and (\ref{102'}), respectively
(see also
(\ref{inhom3}) and (\ref{inhom4'})). 
\end{theorem} 
Note that  the Kolmogorov value $H=1/3$ is covered
by the regimes of Theorem~1 and Theorem~2(i), (iii).

\begin{remark}
As we have seen above, most of the assumptions here 
are motivated by the Gaussian case and we have formulated
them in such a way as to 
allow a significant level of non-Gaussian fluctuation.
In the Gaussian case, we have seen that $\alpha$ in
(\ref{ass})  and (\ref{ass'}) can be chosen to
be any value in $(0,1/4)$.

\end{remark}
\begin{remark}
Both Theorem~1 and 2 can be viewed as a construction
(and the convergence) of approximate solutions
(via Remark~1) to the Gaussian white-noise models
which are widely used in practical applications
\cite{St}, \cite{BO}.

\end{remark}
\commentout{
\begin{remark}
Since both the limiting and pre-limiting equations
preserve the $L^2$-norm of the initial data it
suffices to prove the convergence
in the space $D([0,\infty);L^2_{w}(\IR^{2d}))$ 
where
the weak$-L^2$ topology is used.
\end{remark}
}

\section{Proof of Theorem~1 and 2}
\subsection{Tightness}

In the sequel we will adopt the following notation 
\beq
\label{not}
f_z\equiv f(\lan  W_z^\vas, \theta\ran),\quad
 f'_z\equiv f'(\lan  W_z^\vas, \theta\ran),\quad
f''_z\equiv f''(\lan  W_z^\vas, \theta\ran),\quad 
\quad\forall f\in C^\infty(\IR).
\eeq
Namely, the prime stands for the differentiation w.r.t. the original argument (not
$z$) of $f, f'$ etc. Let $L$ denote the radius of
the ball containing the support of $\cF^{-1}_2\theta$.
Let all the constants $c,c',c_1,c_2,\dots$ etc in
the sequel be independent of $\rho,\eta, \gamma$ and $\ep$
and depend only on $z_0, \theta, \|W_0\|_2$ and $f$.

First we note that since $\cS$ is dense in $L^2(\IR^{2d})$ 
and $\|W^\ep_z\|_2\leq \|W_0\|_2, \forall z>0$, 
 the tightness of  the  family of  $L^2(\IR^{2d})$-valued processes
 $\{ W^\ep, 0<\ep<1\}$ in $ D([0,\infty); L^2_w(\IR^{2d})$
is equivalent to the tightness of
the family in $D([0,\infty);\cS')$ as distribution-valued processes. 
According to \cite{Fougue}, a family 
of processes $\{ W^\ep, 0<\ep<1\} \subset
D([0,\infty); \cS') 
$ is 
tight if and only if  for every test function $\theta\in \cS$ the family
of  processes
$\{\lan  W^\ep, \theta\ran, 0<\ep <1\}
\subset D([0,\infty);\IR)
$  is tight.  With this remark we can now use the tightness 
criterion of \cite{Ku} (Chap. 3, Theorem 4) for finite dimensional
processes, namely,  we will prove:
Firstly,
\beq
\label{trunc}
\lim_{N\to \infty}\limsup_{\ep\to 0}\IP\{\sup_{z<z_0}|\lan  W_z^\ep, \theta\ran|
\geq N\}=0,\quad\forall z_0<\infty.
\eeq
Secondly, for  each $f\in C^\infty(\IR)$  
there is a sequence
$f^\ep_z\in\cD(\cA^\ep)$ such that for each $z_0<\infty$
$\{\cA^\ep f^\ep_z, 0<\ep<1,0<z<z_0\}$ 
is uniformly integrable and
\beq
\lim_{\ep\to 0} \IP\{\sup_{z<z_0} |f^\ep_z-
f(\lan  W_z^\ep, \theta\ran) |\geq \delta\}=0,\quad \forall \delta>0.
\eeq
Then it follows that the laws of
$\{\lan  W^\ep, \theta\ran, 0<\ep <1\}$ are tight in the space
of $D([0,\infty);\IR)$ and hence $\{W^\ep_z\}$ is tight in $
D([0,\infty); L^2_w(\IR^{2d}))$.

Condition (\ref{trunc}) is satisfied because the $L^2$-norm is
preserved. 

We shall construct a test function of the form
$f^\ep_z=f_z+f_{1,z}^\ep+f_{2,z}^\ep+f_{3,z}^\ep.$
First we construct the first perturbation $f_{1,z}^\ep$.
Let
\[
\tilde{V}^\ep_z= \tilde{V}_{z/\ep^2}.
\]
Recall that
\[
\cA^\ep \tilde{V}^\ep_z=-\ep^{-2} V^\ep_z.
\]

Let
\beq
\label{f1}
f_{1,z}^\vas&\equiv &\frac{\ks}{\vas}\int_z^\infty
\, f'_z \lan  W_z^\vas, \mathbb{E}_z^\vas\cV^\ep_s\theta\ran\,ds\\
&=&{\ks}{\vas} f'_z \lan \cF^{-1}_2 W_z^\ep,
\gamma^{-1} \delta_\gamma \int^\infty_z \IE_z[ V^\ep_s]  ds
\cF^{-1}_2
\theta\ran\nn \\
&=&{\ks}{\vas} f'_z \lan \cF^{-1}_2 W_z^\ep,
\gamma^{-1} \delta_\gamma\tilde{V}^\ep_z \cF^{-1}_2
\theta\ran \nn\\
&=&{\ks\vas}
f'_z \lan  W_z^\vas,\cvtil\theta\ran\nn
\eeq
be the 1-st perturbation of $f_z$. 

\begin{prop}\label{prop:2}
$$\lim_{\ep\to 0}\sup_{z<z_0} \mathbb{E}
|f_{1,z}^\vas|=0,\quad
\lim_{\ep\to 0}\sup_{z<z_0} |f_{1,z}^\vas|= 0
\quad \hbox{in probability}$$.
\end{prop}

\begin{proof}
First
\beq
\label{1.2}
\mathbb{E}[|f_{1,z}^\vas|]&\leq &{\vas}\|f'\|_\infty
\| W_0\|_2 
\IE\|\tilde{\cV}^\ep_z\theta\|_2\\
&\leq& c {\vas}\|f'\|_\infty
\| W_0\|_2 
\sup_{|\bx|,|\by|\leq L}
\IE^{1/2}\lt[\gamma^{-1}\delta_\gamma
\tilde{V}^\ep_z(\bx,\by)\rt]^2 \nn\\
&=&O\lt(\ep\eta^{-1}\minrg^{1-H}\rt)
\eeq
which is of the following order of magnitude:
\beq
\label{ta1}
\lt\{\begin{array}{lll}
\ep,& &\hbox{if $\eta,\rho$ held
fixed}\\
\ep, &&\hbox{if $\gamma, \eta$ held fixed}\\
\ep\eta^{-1}, &&\hbox{if $\gamma$ or $ \rho$ held fixed}\\
\ep\minrg^{1-H},&&
\hbox{if
$\eta$ is held fixed}.
\end{array}\rt.
\eeq
and vanishes in the respective regimes.
Secondly, we have 
\beq
\label{1.3}
\sup_{z< z_0} |f_{1,z}^\vas|
& \leq &{\vas}
\|f'\|_\infty  \| W_0\|_2
\sup_{z<z_0}\gamma^{-1}\|\delta_\gamma \tilde{V}^\ep_z
\cF^{-1}_2\theta\|_2\nn\\
&\leq& c
\ep^{1/2}\sup_{|\bx|,|\by|\leq
L}\IE^{1/2}|\gamma^{-1}\delta_\gamma
\tilde{V}^\ep_z(\bx,\by)|^2\nn\\
&=&O\lt(\ep^{1/2}\eta^{-1}\minrg^{1-H}+\ep^{1/2}\eta^{-1}\rho^{2-H}\rt)
\eeq
by (\ref{21}),  with a random constant $c$
possessing  finite moments,  which is of the following order of
magnitude:
\beq
\label{ta2}
\lt\{\begin{array}{lll}
\ep^{1/2},& &\hbox{if $\eta,\rho$ held
fixed}\\
\ep^{1/2}, &&\hbox{if $\gamma, \eta$ held fixed}\\
\ep^{1/2}\eta^{-1}\rho^{2-H}, &&\hbox{if $\gamma$ held
fixed}\\
\ep^{1/2}\eta^{-1}, &&\hbox{if $ \rho$ held
fixed}\\
\ep^{1/2}\minrg^{1-H},&&
\hbox{if
$\eta$ is held fixed}.
\end{array}\rt.
\eeq
which vanishes in the respective regimes.
  The
right side of (\ref{1.3}) now converges to zero
in probability by a simple application
of Chebyshev's inequality and  assumption (\ref{ass}).
\end{proof}

A straightforward calculation yields
\beqn
 \cA^\vas f_1^\vas &=&-
  {\ks\ep}
f'_z\lan\wepz,\lt[\frac{\bp}{\ks}\cdot\nabla+\frac{\ks}{\ep}\cv\rt]\cvtil 
  \theta\ran-
  \frac{\ks}{\ep}f'_z\lan \wepz,\cv\theta\ran
  +
 {\ks\ep} f''_z\lan \wepz, \cA^\ep \theta\ran\lan \wepz, 
    \cvtil\theta\ran
\nn
\eeqn
where $\cA^\ep\theta$ denotes 
\[ \cA^\vas
\theta
=-\frac{1}{\ks}
{\bp}\cdot \nabla_\bx\theta - \frac{\ks}{\vas} 
\cv\theta
\]
cf. (\ref{gen}).
Hence
\beq
\label{25'}
\cA^\vas \lt[f_z+f^\ep_{1,z}\rt]
&=&\frac{1}{\ks}f'_z\lan \wepz, \pdgx \theta\ran
+{\ks^2}f'_z\lan \wepz, \cv\cvtil\theta\ran
+{\ks^2}f''_z\lan\wepz,\cv\theta\ran\lan\wepz,\cvtil\theta\ran
\nn\\
&&\quad
+{\ep}\lt[f'_z\lan\wepz,\pdgx\cvtil\theta\ran
+f''_z\lan\wepz,\pdgx\theta\ran\lan
\wepz,\cvtil\theta\ran\rt]\nn\\
&=&A_1^\vas(z)+A_2^\vas(z)+A_3^\vas(z)+A_4^\vas(z)
\nonumber
\eeq
where $A_2^\vas(z)$ and $A_3^\vas(z)$ are the coupling terms.

\begin{prop}
\label{prop:1}
$$\lim_{\ep\to
0}\sup_{z<z_0}\IE|A^\ep_4(z)|=0$$.
\end{prop}
\begin{proof} By Lemma~2 we have
 \beq
 |A^\ep_4|
  &\leq& {\vas}
  \|f''\|_\infty\| W_0\|^2_2\lt[
  \|\pdgx\theta\|_2\|\cvtil\theta\|_2+
  \|\pdgx(\cvtil\theta)\|_2\rt]\nn
 \label{1.10}
  \eeq
which is of the following order of magnitude
\beq
\label{ta3}
\lt\{
\begin{array}{lll}
\ep,&&\hbox{if $\eta,\rho$ held fixed}\\
\ep\rho^{2-H},&&\hbox{if
$\eta,\gamma$ held fixed} \\
\ep\eta^{-1},&&
\hbox{if $\rho$ is held fixed}\\
\ep(\minrg^{1-H}+\rho^{2-H}),&&\hbox{if
$\eta$ held fixed}
\end{array}\rt.
\eeq
and vanishes in the respective regimes.
\end{proof}

We introduce the next perturbations $f_{2,z}^\ep,
f_{3,z}^\ep$. Let
\bea
\label{40.3}
A_2^{(1)}(\phi) &\equiv&\int\phi(\bx,\bp)
\cQ_1(\theta\otimes\theta)(\bx,\bp,\by,\bq)\phi(\by,\bq)\,d\bx d\bp\,d\by d\bq\\
A_1^{(1)}(\phi)&\equiv&\int\cQ'_1\theta(\bx,\bp)\phi(\bx,\bp)\,\,d\bx d\bp
\label{41.2}
\eea
where 
\beq
    \cQ_1(\theta\otimes\theta)(\bx,\bp,\by,\bq)&=&
    \IE\lt[\cv\theta(\bx,\bp)\cvtil\theta(\by,\bq)\rt]
\eeq
and
\[
\cQ'_1\theta(\bx,\bp)=\IE\lt[\cv\cvtil\theta (\bx,\bp)\rt]
\]
where the operator $\cvtil$ is defined as in (\ref{cvtil}).
Note that $\cQ_1\theta$ and $\cQ'_1\theta$ are $O(1)$ terms
because  of (\ref{19}).

Clearly, we have
\beq
A_2^{(1)}(\phi)=\mathbb{E}\lt[\lan\phi, \cv\theta\ran
\lan\phi, \cvtil\theta\ran\rt].
\eeq

Define
\begin{align*}
f_{2,z}^\vas &\equiv
{\ks^2}f''_z\int_z^\infty \mathbb{E}_z^\vas
\lt[\lan\tvas, \cV^\ep_s\theta\ran\lan\tvas,\tilde{\cV}^\ep_s\theta\ran -A^{(1)}_2(\tvas)\rt]\,ds
\\
f_{3,z}^\vas &\equiv {\ks^2}f'_z\int_z^\infty
\mathbb{E}_z^\vas
\lt[\lan\tvas, \cV^\ep_s\tilde{\cV}^\ep_s\theta\ran-A^{(1)}_3(\tvas)\rt]\,ds.
\end{align*}

Let
\[
\cQ_2(\theta\otimes\theta)(\bx,\bp,\by,\bq)\equiv
\IE\lt[\veptil\theta(\bx,\bp)\veptil\theta(\by,\bq)\rt]
\]
and \[
\cQ'_2\theta(\bx,\bp)=\IE\lt[\cvtil\cvtil\theta (\bx,\bp)\rt].
\]
Let
\bea
A_2^{(2)}(\phi) &\equiv&\int\phi(\bx,\bp)
\cQ_2(\theta\otimes\theta)(\bx,\bp,\by,\bq)\phi(\by,\bq)\,\,d\bx d\bp\,d\by
d\bq\\
A_1^{(2)}(\phi)&\equiv&\int\cQ_2'\theta(\bx,\bp)\phi(\bx,\bp)\,\,d\bx\,d\bp
\eea
we then have
\beq
f_{2,z}^\vas &=&\frac{\ep^2\ks^2}{2}f''_z
\lt[\lan\psiep,\cvtil\theta\ran^2-A^{(2)}_2(\psiep)\rt]\\
f_{3,z}^\vas &=&
\frac{\ep^2\ks^2}{2}f'_z\lt[\lan\psiep,\cvtil\cvtil\theta\ran-
A^{(2)}_3(\psiep)\rt].
\eeq

\begin{prop}\label{prop:4}
$$ \lim_{\ep\to 0}\sup_{z<z_0}
\mathbb{E}|f_{j,z}^\vas|=0,\quad \lim_{\ep\to
0}\sup_{z<z_0} |f_{j,z}^\vas|=0, \quad j=2,3. $$
\end{prop}
\begin{proof}
We have the bounds
\beqn
\sup_{z<z_0}\IE|f_{2,z}^\vas|&\leq&
\sup_{z<z_0}{\vas^{2}\ks^2}\|f''\|_\infty\lt[
\| W_0\|_2^2\IE\|\cvtil\theta\|_2^2
+\IE[A^{(2)}_2(\psiep)]\rt]\\
\sup_{z<z_0}
\IE|f_{3,z}^\vas|&\leq&
\sup_{z<z_0}
{\vas^{2}\ks^2}\|f'\|_\infty
\lt[\| W_0\|_2\IE \|\cvtil\cvtil\theta\|_2
+\IE[A_1^{(2)}(\psiep)]\rt].
\eeqn
The first term can be estimated as in (\ref{ta1});
the second term can by estimated as in (\ref{ta1}) by using the
 (\ref{gauss3}).

As for estimating $\sup_{z<z_0} |f_{j,z}^\vas|, j=2,3$, we
have
\beqn
\sup_{z<z_0}|f_{2,z}^\vas|&\leq&
\sup_{z<z_0}{\vas^2\ks^2}\|f''\|_\infty\lt[
\| W_0\|_2^2\|\cvtil\theta\|_2^2
+A^{(2)}_2(\psiep)\rt]\\
\sup_{z<z_0}
|f_{3,z}^\vas|&\leq&
\sup_{z<z_0}
{\vas^2\ks^2}\|f'\|_\infty
\lt[\| W_0\|_2\|\cvtil\cvtil\theta\|_2
+A_1^{(2)}(\psiep)\rt].
\eeqn
Using the assumption  (\ref{21})   we can estimate the right
side of the above as in (\ref{ta2}). 

\end{proof}

We have
\beqn
\cA^\vas f_{2,z}^\vas&=&{\ks^2}
f''_z\left[-
\lan\tvas, \cv\theta\ran\lan\tvas,\cvtil\theta\ran + A^{(1)}_2(\tvas)\right]
+ R_2^\vas(z)\\
\cA^\vas f_{3,z}^\vas&=&{\ks^2}
f'_z\left[-\lan
\tvas,\cv\cvtil\theta\ran + A^{(1)}_3(\tvas)\right]+
R_3^\vas(z)
\eeqn
with
\beq
\nn
R_2^\vas(z)& =&\ep^2\frac{\ks^2}{2}{f'''_z}\left[
\frac{1}{\ks}
\lan\psiep,\pdgx\theta\ran+\frac{\ks}{\ep}\lan\psiep,\cv\theta\ran\rt]
\lt[\lan\psiep,\cvtil\theta\ran^2-A_2^{(2)}(\psiep)\rt]\\
\nn
&& +
\ep^2{\ks^2}f''_z\lan\psiep,\cvtil\theta\ran\lt[
\frac{1}{\ks}\lan\psiep,\pdgx(\cvtil\theta)\ran+\frac{\ks}{\ep}
\lan\psiep,\cv\cvtil\theta\ran\rt]\\
&&-\ep^2{\ks^2}f'_z\lt[\frac{1}{\ks}\lan\psiep,\pdgx(G_\theta^{(2)}\psiep)\ran+
\frac{\ks}{\ep}\lan\psiep,\cv G_\theta^{(2)}\psiep\ran\rt]
\label{r2}
\eeq
where 
$G_\theta^{(2)}$ denotes the operator
\[
G_\theta^{(2)}\phi\equiv \int \cQ_2(\theta\otimes\theta)(\bx,\bp,\by,\bq)\phi(\by,\bq)
\,d\by d\bq.
\]
Similarly
\beq
R^\ep_3(z)&=&\ep^2{\ks^2}
f'_z\left[\frac{1}{\ks}\lan\psiep,\pdgx(\cvtil\cvtil\theta)\ran+\frac{\ks}{\ep}
\lan\psiep,\cv\cvtil\cvtil\theta\ran\rt]\nn\\
&&+\ep^2\frac{\ks^2}{2}
f''_z\left[
\frac{1}{\ks}\lan\psiep,\pdgx\theta\ran+\frac{\ks}{\ep}\lan\psiep,\cv\theta\ran\rt]
\lt[\lan\psiep,\cvtil\cvtil\theta\ran-A_1^{(2)}(\psiep)\rt]\nn\\
&&-\ep^2{\ks^2}f'_z\lt[
\frac{1}{\ks}\lan\psiep,\pdgx(\cQ_2'\theta)\ran+\frac{\ks}{\ep}
\lan\psiep,\cv\cQ_2'\theta\ran\rt].
\label{r3}
\eeq

\begin{prop}\label{prop:5}
\[
\lim_{\ep\to 0}\sup_{z<z_0} \mathbb{E} |R_2^\vas(z)|=0,\quad \lim_{\ep \to 0}
\sup_{z<z_0} \mathbb{E}
|R_3^\vas(z)|=0.
\]
\end{prop}
\begin{proof}
Part of the argument is analogous to that  given for
Proposition~\ref{prop:4}.  The additional estimates that we need to consider
are  the following. 

In $R_2^\ep$  (\ref{r2}):
\beqn
\label{r2.1}\nn
\lefteqn{\sup_{z<z_0}\ep^2
\IE\lt|\lan\psiep,\pdgx(G_\theta^{(2)}\psiep)\ran\rt|}\\ &\leq &
c\ep^2\gamma^{-2}\|W_0\|_2 \IE\lt\{\lt\|\nabla_\by\cdot\nabla_\bx
  \cF^{-1}_2\theta (\bx,
\by)\right. \right.\nn\\
&&\quad\quad \quad\quad\quad\quad \quad\lt. \lt.
\times\int\IE\lt[\delta_\gamma
\tilde{V}^\ep_z(\bx,\by)\delta_\gamma
\tilde{V}^\ep_z (\bx',\by')\rt]\cF^{-1}_2\theta
(\bx',\by')\cF^{-1}_2
\wepz(\bx',\by')d\bx' d\by'\rt\|_2\rt\}\nn\\
&\leq & c\ep^2\gamma^{-2}\|W_0\|_2 \IE\lt\{\lt\|\nabla_\by\cdot\nabla_\bx
  \cF^{-1}_2\theta (\bx,
\by)\IE\lt[\delta_\gamma
\tilde{V}^\ep_z(\bx,\by)\rt]^2\int \lt|\cF^{-1}_2\theta (\bx',\by')\cF^{-1}_2
\wepz(\bx',\by')\rt| d\bx' d\by'\rt\|_2\rt\}\nn\\
&\leq & c\ep^2\gamma^{-2}\|W_0\|_2 \lt\|\nabla_\by\cdot\nabla_\bx
  \cF^{-1}_2\theta \IE\lt[\delta_\gamma
\tilde{V}^\ep_z\rt]^2\rt\|_2\IE\lt\|\cF^{-1}_2\theta \cF^{-1}_2
\wepz\rt\|_2\nn\\
&\leq & c\ep^2\gamma^{-2}\|\theta\|_2\|W_0\|_2^2 \lt\|\nabla_\by\cdot\nabla_\bx
  \cF^{-1}_2\theta \IE\lt[\delta_\gamma
\tilde{V}^\ep_z\rt]^2\rt\|_2\nn\\
  &\leq&
  c\|\theta\|_2\|W_0\|_2^2\ep^2\gamma^{-1}\lt\|
    [\cF^{-1}_2\nabla_\bx\cdot\nabla_\bx\theta](\bx,\by)
    \IE\lt[\delta_\gamma
    \tilde{V}^\ep_z\rt]^2(\by)\rt\|_2\nn\\
    &&+c\|\theta\|_2\|W_0\|_2^2\ep^2\gamma^{-2}
    \lt\|
      [\cF^{-1}_2\nabla_\bx\theta](\bx,\by)\cdot
      \nabla_\by\IE\lt[\delta_\gamma
      \tilde{V}^\ep_z \rt]^2(\by)\rt\|_2\nn\\
      &\leq&
      c\|\theta\|_2\|W_0\|_2^2\ep^2\gamma^{-1}
      \sup_{|\by|\leq L}\IE\lt[\delta_\gamma
      \tilde{V}^\ep_z\rt]^2(\by)+c\|\theta\|_2\|W_0\|_2^2\ep^2\gamma^{-2}
       \sup_{|\by|\leq L}
       \lt|\nabla_\by\IE\lt[\delta_\gamma
       \tilde{V}^\ep_z \rt]^2(\by)\rt|\\
&\leq&O\lt(\ep^2\eta^{-2} \gamma\minrg^{2-2H}
+\ep^2\eta^{-2}\rho^{1-H}\minrg^{1-H}\rt)
       \eeqn
     by Lemma~\ref{newlemma2}  where  $L$ is the radius of
the ball containing
         the support of $\theta$. 
Further delineation yields
\beqn
\sup_{z<z_0}\ep^2
\IE\lt|\lan\psiep,\pdgx(G_\theta^{(2)}\psiep)\ran\rt|
=\lt\{\begin{array}{ll}
\ep^2& \hbox{if $\eta, \rho$ held fixed}\\
\ep^2\rho^{1-H} &\hbox{if $\eta, \gamma$ held fixed}\\
\ep^2\eta^{-2}\rho^{1-H} &\hbox{if $\gamma$ held fixed}\\
\ep^2\eta^{-2}&\hbox{if $\rho$ held fixed}\\
\ep^2 \rho^{1-H}\minrg^{1-H}&\hbox{if $\eta$ held fixed}.
\end{array}
\rt.
\eeqn
          
Consider the next term:
\beq
\label{r2.2}\nn
\lefteqn{\sup_{z<z_0}\ep \IE\lt|\lan\psiep,\cv G_\theta^{(2)}\psiep\ran\rt|}\\
&\leq& c\ep^2\gamma^{-3}\|W_0\|_2 \IE\lt\{\lt\|\delta_\gamma
V^\ep_z(\bx,\by)
  \cF^{-1}_2\theta (\bx,
\by)\rt.\rt. \nn\\
&&\lt.\lt.\quad\quad\quad\quad\quad\quad \quad \times
\int\IE\lt[\delta_\gamma
\tilde{V}^\ep_z(\bx,\by)\delta_\gamma
\tilde{V}^\ep_z (\bx',\by')\rt]\cF^{-1}_2\theta
(\bx',\by')\cF^{-1}_2
\wepz(\bx',\by')d\bx' d\by'\rt\|_2\rt\}\nn\\
&\leq & c\ep^2\gamma^{-3}\|W_0\|_2  \IE\lt\{\lt\|\delta_\gamma
V^\ep_z(\bx,\by)
  \cF^{-1}_2\theta (\bx,
\by)\IE\lt[\delta_\gamma
\tilde{V}^\ep_z(\bx,\by)\rt]^2\right.\rt.\nn\\
\nn&&\hspace{4cm}\lt.\lt.\times\int \lt|\cF^{-1}_2\theta
(\bx',\by')\cF^{-1}_2
\wepz(\bx',\by')\rt| d\bx' d\by'\rt\|_2\rt\}\nn\\
&\leq & c\ep^2\gamma^{-3}\|\theta\|_2\|W_0\|^2_2 \IE\lt\|\delta_\gamma
V^\ep_z(\bx,\by)
  \cF^{-1}_2\theta \IE\lt[\delta_\gamma
\tilde{V}^\ep_z\rt]^2\rt\|_2\nn\\
&\leq&O\lt(\ep^2\eta^{-2}\minrg^{3-3H}\rt)\nn
\eeq
by Corollary~\ref{cor3}.

In $R_3^\ep$ (\ref{r3}):
\beq
\label{r3.1}
\sup_{z<z_0}\ep  \IE\lt|\lan\psiep,\cv\cvtil\cvtil\theta\ran\rt| &\leq&
\ep \|W_0\|_2 \sup_{z<z_0}\sqrt{\IE \lt\|\cv\cvtil\cvtil\theta\rt\|^2_2}\nn\\
&=& O\lt(\ep\gamma^{-3}\sup_{|\by|\leq L}\IE\lt|\delta_\gamma
\tilde{V}^\ep_z\rt|^2(\by)
\IE^{1/2}\lt|\delta_\gamma V^\ep_z\rt|^2(\by)\rt)\nn\\
&=&O\lt(\ep\eta^{-2}\minrg^{3-3H}\rt),
\nn
\eeq
by (\ref{gauss4}) and Lemma~\ref{newlemma2}, which can be
further delineated in the various regimes as
\beqn
\lt\{\begin{array}{ll}
\ep&\hbox{if $\rho$ and $\eta$ held fixed}\\
\ep&\hbox{if $\gamma$ and $\eta$ held fixed}\\
\ep\eta^{-2} &\hbox{if $\gamma$ or $\rho$ held fixed}\\
\ep\minrg^{3-3H}&\hbox{if $\eta$ held fixed};
\end{array}
\rt.
\eeqn
\beq
\label{r3.2}
\ep^2
\IE\lt|\lan\psiep,\pdgx(\cQ_2'\theta)\ran\rt|&\leq&
\ep^2
\sqrt{\IE\lt|\lan\psiep,\pdgx(\cQ_2'\theta)\ran\rt|^2}\\
&\leq & c\ep^2\gamma^{-2}\|W_0\|_2 \lt\|\nabla_\by\cdot\nabla_\bx
  \IE\lt[\delta_\gamma
\tilde{V}^\ep_z(\bx,\by)\rt]^2 \cF^{-1}_2\theta (\bx,
\by)\rt\|_2\nn\\ &=&O\lt(\ep^2\gamma^{-2}\IE_{|\by|\leq
L}\lt|\nabla_\by\IE\lt[
\delta_\gamma \tilde{V}^\ep_z\rt]^2(\by)\rt|\rt)\nn\\
&=&O\lt(\ep^2\eta^{-2}\rho^{1-H}\minrg^{1-H}\rt)
\nn
\eeq
which in the various regimes has the following
order of magnitude
\beqn
\lt\{\begin{array}{ll}
\ep^2&\hbox{if $\rho$ and $\eta$ held fixed}\\
\ep^2\rho^{1-H}&\hbox{if $\gamma$ and $\eta$ held fixed}\\
\ep^2\eta^{-2}\rho^{1-H} &\hbox{if $\gamma$ held fixed}\\
\ep^2\eta^{-2} &\hbox{if $\rho$ held fixed}\\
\ep^2\rho^{1-H}\minrg^{1-H}&\hbox{if $\eta$ held fixed};
\end{array}
\rt.
\eeqn

\beq
\label{r3.3}
{\ep}
\IE\lt|\lan\psiep,\cv\cQ_2'\theta\ran\rt| &\leq &
{\ep}
\sqrt{\IE\lt|\lan\psiep,\cv\cQ_2'\theta\ran\rt|^2}\\
&\leq & c\ep^2\gamma^{-3}\|W_0\|_2 \IE \lt\|\delta_\gamma
V^\ep_z(\bx,\by)
  \IE\lt[\delta_\gamma
\tilde{V}^\ep_z(\bx,\by)\rt]^2 \cF^{-1}_2\theta (\bx,
\by)\rt\|_2\nn\\ &=& O\lt(\ep^2\gamma^{-3}\sup_{|\by|\leq
L}\IE\lt|\delta_\gamma \tilde{V}^\ep_z\rt|^2(\by)
\IE^{1/2}\lt|\delta_\gamma V^\ep_z\rt|^2(\by)\rt)\nn\\
&=&O\lt(\ep^2\eta^{-2}\minrg^{3-3H}\rt)
\nn
\eeq
by Lemma~\ref{newlemma2}.
\end{proof}

Consider the test function $f^\ep_z=f_z+f_{1,z}^\ep
+f_{2,z}^\ep+f_{3,z}^\ep$. 
We have
\beq
\label{2.67}
\label{A2}
\cA^\ep f^\ep_z
&=&
\frac{1}{\ks}f'_z\lan\psiep,\pdgx\theta\ran+{\ks^2}f''_z
A_2^{(1)}(\psiep)+{\ks^2}f'A_1^{(1)}(\psiep)
+R_2^\ep(z)+R_3^\ep(z)+A_4^\ep(z).
\eeq
Set
\beq
\label{remainder}
R^\vas(z) = R_1^\vas(z) + R_2^\vas(z) + R_3^\vas(z),\quad\hbox{with}\,\,
R_1^\vas(z)=A_4^\ep(z).
\eeq
It follows from Propositions~\ref{prop:1} and \ref{prop:5} that
\[
\lim_{\ep \to 0}\sup_{z<z_0}\IE|R^\ep(z)|=0.
\]

For the tightness it remains to show 

\begin{prop}
\label{prop:7}
$\{\cA^\ep f^\ep_z\}$ are uniformly integrable.
\end{prop}
\begin{proof}
We shall prove that each term in the expression (\ref{2.67}) is
uniformly integrable. We only need to   be concerned with terms
in $R^\ep(z)$ since other terms are obviously uniformly
integrable
because $W^\ep_z$ is uniformly  bounded in  the square
norm.
But since the previous estimates
establish the uniform boundedness of the second moments of the
corresponding terms, the uniform integrability of the terms follow.

\end{proof}

\subsection{Identification of the limit}
Our strategy  is to show directly
that in passing to the weak limit  the limiting process
solves the martingale problem  formulated in Section~2.1.
The uniqueness
of the martingale solution mentioned in Section~\ref{npt} then
identifies the limiting process as the  unique $L^2(\IR^{2d})$-valued
solution  to  the initial value problem of
the stochastic PDE (\ref{ito}).

\commentout{
With this remark and the tightness result we can use another result 
 in \cite{Ku} (Chapter 3, Theorem 2) to identify the limit.
Let $\bar{\cA}$ be a diffusion or jump diffusion operator such that
there is a unique solution $\omega_z$ in the space 
$D([0,\infty);L^2_{w}(\IR^{2d}))$
 such that 
\beq
\label{38}
f(\omega_z)-\int^z_0\bar{\cA} f(\omega_s)\,ds
\eeq
is a martingale. 
We shall show that for each $f\in C^\infty(\IR)$ there exists
$f_z^\ep\in \cD(\cA^\ep)$ such that
\beq
\label{38.2}
\sup_{z<z_0,\ep}\IE|f^\ep_z-f(\lan 
W^\ep_z,\theta\ran)|&<&\infty\\
\label{39.2}
\lim_{\ep\to 0}\IE|f^\ep_z-f(\lan 
W^\ep_z,\theta\ran)|&=&0,\quad \forall z<z_0\\
\label{40.2}
\sup_{z<z_0,\ep}\IE|\cA^\ep f^\ep_z-\bar{\cA} f(\lan 
W^\ep_z,\theta\ran)|&<&\infty\\
\lim_{\ep\to 0}\IE|\cA^\ep f^\ep_z-\bar{\cA} f(\lan 
W^\ep_z,\theta\ran)|&=&0,\quad
\forall z<z_0.
\label{42.2}
\eeq
Then the aforementioned theorem implies that any tight  family of processes
$\lan  W^\ep_z,\theta\ran$ converge
in law to the unique process generated by $\bar{\cA} $.
}

Recall that for any $C^2$-function $f$
\beq
M_z^\vas(\theta)&=&f^\ep_z-\int^z_0 \cA^\ep
f^\ep_s\,ds\label{77}\\ &=&
f_z+f_1^\vas(z)+f_2^\vas(z)+f_3^\vas(z) -
\int_0^z\frac{1}{\ks}f'_z\lan\tvas,\pdgx\theta\ran\,ds\nn\\
&& - \int_0^z{\ks^2}\left[f''_s
A_2^{(1)}( W_s^\vas)+f'_s A_1^{(1)}( W_s^\vas)\right]\,ds -\int_0^z
R^\vas(s)\,ds\nn
\eeq
is a martingale. The martingale property implies that for any finite
sequence
$ 0<z_1<z_2<z_3<...<z_n \leq z $,  $C^2$-function $f$ and 
bounded continuous  function $h$ with compact support,
we have 
\beq
\label{78}
\IE\lt\{h \lt(\lan W^\ep_{z_1}, \theta\ran,
\lan W^\ep_{z_2},\theta\ran,...,\lan W^\ep_{z_n},\theta\ran\rt)
\lt[M^\ep_{z+s}(\theta)-M^\ep_{z}(\theta)\rt]\rt\}&=&0,\\
\quad \forall
s>0,\quad  z_1\leq z_2\leq\cdots\leq z_n\leq z.&&\nn
\eeq
Let
\[
\bar{\cA} f_z\equiv f'_s\left[\frac{1}{\ks}\lan
 W_z,\pdgx\theta\ran+{\ks^2}\bar{A}_1( W_z)\right]
 + {\ks^2}f''_z\bar{A}_2( W_z)
\]
where
\beq
\label{doom}
\bar{A}_2(\theta)=\lim_{{\rho}\to \infty}{A}_2^{(1)}(\theta)
=\overline{Q}(\theta\otimes\theta),
\quad \bar{A}_1(\theta)=\lim_{{\rho}\to\infty}{A}_1^{(1)}(\theta)=
\overline{Q}_0(\theta)
\eeq
as given in  (\ref{101}) and (\ref{102}),
respectively. For $\rho\to\infty, \gamma\to 0$ as $\ep\to 0$ the limits
in (\ref{doom}) are not well-defined unless $H\in (0,1/2)$
in the worst case scenario allowed by (\ref{power}). 
Likewise, the convergence does not hold for $H\in [1/2,1)$
when $\eta\to 0$ in the worst case scenario allowed by (\ref{power}).

For each possible limit process in $D([0, \infty);L^2_w(\IR^{2d}))$
there is at most a countable set of discontinuous points
with a positive probability and we consider all the finite
set  $\{z_1, ...,z_n\}$ in
(\ref{78}) to be outside of the set of discontinuity.

In view of the results of
Propositions~\ref{prop:2}, \ref{prop:1}, \ref{prop:4}, 
\ref{prop:5}  we see
that
$f^\ep_z$ and
$\cA^\ep f^\ep_z$ in (\ref{77}) can be replaced by $f_z$ and
$\bar{\cA}f_z$, respectively, modulo an error that vanishes as
$\ep\to 0$.
With this and the tightness of $\{W^\ep_z\}$ 
we can pass to the limit
$\ep\to 0$ in  (\ref{78}).  
We see that the limiting process satisfies the 
martingale property that
\beqn
\IE\lt\{h \lt(\lan W_{z_1}, \theta\ran,
\lan W_{z_2},\theta\ran,...,\lan W_{z_n},\theta\ran\rt)
\lt[M_{z+s}(\theta)-M_{z}(\theta)\rt]\rt\}=0,\quad \forall s>0.
\eeqn
where
\begin{equation}
M_z(\theta)=f_z-\int_0^z \bar{\cA}f_s
\,ds.
 \label{37}
\end{equation}
Then it follows that
\[
\IE\lt[ M_{z+s}(\theta)-M_z(\theta)|W_u, u\leq z\rt]=0,\quad
\forall z,s>0
\]
which proves that $M_z(\theta)$ is a martingale.

Note that  $\lan\tvas,\theta\ran$ is uniformly bounded:
\[
\lt|\lan\tvas,\theta\ran\rt|\leq \| W_0\|_2{\|\theta\|}_2
\]
so we have the convergence of the second moment
\[
\lim_{\ep\to 0}
\IE\left\{
{\lan\tvas,\theta\ran}^2\right\}=\mathbb{E}\left\{ {\lan
 W_z,\theta\ran}^2\right\}.
\]
Using $f(r) =r$ and $r^2$ in (\ref{37}) we see that
\[
M_z^{(1)}(\theta)=\lan  W_z,\theta\ran -
\int_0^z \left[\frac{1}{\ks}\lan  W_s,\pdgx\theta\ran -
{\ks^2}\bar{A}_3( W_s)\right]\,ds \]
is a martingale with the
quadratic variation
\[
\left[M^{(1)}(\theta),M^{(1)}(\theta)\right]_z=
{\ks^2}\int_0^z\bar{A}_2( W_s)\,ds
={\ks^2}\int^z_0\lan W_s, \overline{\cK}_{\theta}W_s\ran\,ds
\]
where $\overline{\cK}_{\theta}$ is defined as in (\ref{r10}).

\begin{appendix}
\section{Mixing coefficients and moment estimates for $\tilde{V}_z$}
Let $\cF_z $ and $\cF^+_z$  be the sigma-algebras generated by
$\{V_s:  \forall s\leq z\}$ and $\{V_s: \forall s\geq  z\}$, respectively.

 Consider the strong mixing coefficient
\beqn
\alpha(t)&=&\sup_{A\in \cF^+_{z+t}}\sup_{B\in \cF_z}
|P(AB)-P(A)P(B)|\\
&=&\frac{1}{2}\sup_{A\in \cF_{z+t}}\IE\lt[ |P(A|\cF_z)-P(A)|\rt]
\eeqn
which can be used to bound the first order moment:
\[
\IE\lt[ |\IE\lt[V_s|\cF_z\rt]|\rt]\leq 8
\alpha(s-z)^{1/p}\lt[\IE|V_s|^q\rt]^{1/q},\quad \forall s>z,\quad
p^{-1}+q^{-1}=1
\]
(\cite{EK}, Corollary 2.4). Hence
 the integrability
of $\alpha(t)$ implies that $\tilde{V}_z$  has a finite first order
moment.

 To bound the higher order moments of
$\tilde{V}_z$ one can  consider, for example, 
the general $L^p$-mixing coefficients
\beqn
\phi_p(t)
&=&\sup_{A\in \cF_{z+t}}\IE^{1/p}\lt[ |P(A|\cF_z)-P(A)|^p\rt],\quad
p\in [1,\infty)\\
&=& \sup_{h\in L^p(P,\cF^+_{z+t})}\sup_{g\in L^q(P,
\cF_z)\atop
\IE g^q=1, \IE g=0} \IE[ hg],\quad p^{-1}+q^{-1}=1,\quad p\in
[1,\infty)
\eeqn
We note that $\alpha(t)=\phi_1(t)$ and for $p=\infty$
\beqn
\phi_\infty(t)&=&\sup_{A\in \cF^+_{t+z}}\sup_{B\in
\cF_z\atop P(B)>0}|P(A|B)-P(A)|, \quad \forall t\geq 0\\
&=&\sup_{A\in
\cF^+_{t+z}}\hbox{ess-sup}_\omega|P(A|\cF_z)-P(A)|\\
&\equiv& \phi(t)
\eeqn
 is called
the uniform mixing coefficient \cite{EK}.   In terms of $\phi_p$
one has the following estimate
\beq
|\IE\lt[h_1 h_2\rt]-\IE[h_1]\IE[h_2]|
\leq 2^{\min{(q,2)}}\phi_p(t)^{1/u}
\IE^{1/(vp)}[h_2^{vp}]\IE^{1/q}[h^q_1]
\eeq
for $u, v, p, q\in [1,\infty], u^{-1}+v^{-1}=1, p^{-1}+q^{-1}=1$
and real-valued $h_1\in L^q(\Omega, \cF_z, P), h_2
\in L^{vp}(\Omega, \cF_{z+t}^+, P)$ (see  \cite{EK},  Proposition~2.2). 
In particular, for $q>2, v=q/p$,
\beq
|\IE\lt[h_1 h_2\rt]-\IE[h_1]\IE[h_2]|
\leq 4\phi_p(t)^{(q-p)/q}
\IE^{1/q}[h_2^{q}]\IE^{1/q}[h^q_1],\quad p^{-1}+q^{-1}=1
\eeq
by which, along with the  H\"older inequality, we can bound the
second moment of
$\tilde{V}_z$ as follows:
First we observe that for $s, \tau\geq z$ and $h_1= \IE_z(V_s),$ 
$h_2= V_\tau$  
\beqn
\IE\lt[ \IE_z[V_s(\bx)]\IE_z[V_\tau(\bx)]\rt]=
\IE\lt[\IE_z[V_s(\bx)] V_\tau(\bx)\rt]
\leq 4\phi_p(\tau-z)^{(q-p)/q} \IE^{1/q}[V_z^{q}]
\IE^{1/q}[\IE^{q}_z\lt[V_s\rt]].
\eeqn
By  setting $s=\tau$ first and the Cauchy-Schwartz inequality we
have
\beqn
\IE\lt[\IE^2_z[V_s]\rt]&\leq& 4\phi_p(s-z)^{(q-p)/q}
\IE^{2/q}[V_z^{q}]\\
\IE\lt[ \IE_z[V_s(\bx)]\IE_z[V_\tau(\bx)]\rt]
&\leq &4\phi_p(s-z)^{(q-p)/(2q)} \phi_p(\tau-z)^{(q-p)/(2q)}
\IE^{2/q}[V_z^{q}], \quad s, \tau \geq z.
\eeqn
Hence
\beqn
\IE[\tilde{V}_z^2]
&\leq&
2\int^\infty_z \int^\infty_z
\IE\lt[\IE_z[V_\tau]\IE_z\lt[V_s\rt]\rt]ds d\tau+2\int^\infty_0
\int^\infty_0 \IE\lt[\IE_0[V_\tau]\IE_0\lt[V_s\rt]\rt]ds d\tau\\ 
&\leq& 8\IE^{2/q}[V_z^{q}]
\lt(\int^\infty_0\phi_p(t)^{(q-p)/(2q)}dt\rt)^2\\
&\leq &8 \IE^{1/3}[V_z^6]
\lt(\int^\infty_0\phi_{6/5}^{2/5}(t) dt\rt)^2
\eeqn
 which is
finite if
$\phi_{6/5}^{2/5}(t)
$ is integrable (if $V_z$ is assumed to have a finite  6-th order
moment).

When $V_z$ is almost surely bounded, the preceding calculation
with $p=1, q=\infty$ becomes
\beqn
\IE[\tilde{V}_z^2]
&\leq &8  \lim_{q \to \infty}\IE^{1/q}[V_z^q]
\lt(\int^\infty_0\phi_{1}^{1/2}(t) dt\rt)^2
\eeqn
which is finite when $\phi_1^{1/2}(t)$ is integrable.

In order to bound higher order moments
in the non-Gaussian case,  one
can assume the integrability of the uniform  mixing
coefficient $\phi(t)\equiv\phi_\infty(t)$.
Then we have
\[
|P(A|\cF_z)-P(A)|\leq \phi(s-z),\quad\forall A\in \cF_{s}, \quad
s\geq z
\]
and for $p\in [1,\infty), p^{-1}+q^{-1}=1$
\beq
\lt|\IE\lt[V_s|\cF_z\rt]\rt|\leq
2^{1/p}\phi^{1/p}(s-z)\lt[
\IE\lt[V_s^q|\cF_z\rt]
+\IE\lt[V_s^q\rt]\rt]^{1/q}
\label{umix}
\eeq
(cf. \cite{EK}, Proposition 2.6). 
Using (\ref{umix}) and the H\"oloder
inequality repeatedly we obtain
\beq
\IE\lt\{\int^\infty_z
\IE\lt[V_s|\cF_z\rt]ds\rt\}^p&\leq &
c \lt[\int^\infty_0 \phi(s)ds\rt]^p
\IE[V_s^p].
\eeq
Hence the integrability of $\phi(t)$
implies  that
$\tilde{V}_z$ given by (\ref{new}) has a finite moment
of any order $p<\infty$ if  $V_z$ has a finite moment
of  order $p$.

In summary we have
\begin{proposition}
\begin{itemize}
\item[(i)] Assume that 
$\IE[V_z^p]<\infty, p\in [1,\infty)$. 
If the uniform ($L^\infty$-) mixing
coefficient 
$\phi_\infty(t)$ of $V_z$ is integrable
then
$\tilde{V}_z$ has finite moments of order
$p$.
\item[(ii)] Assume that $\IE[V_z^6]<\infty$. If  the
$2/5$-power of the $L^{6/5}$-mixing coefficient 
$\phi_{6/5}(t)$ is integrable, then
$\tilde{V}_z$ has  finite second
moment.
\item[(iii)] Assume $V_z$ is almost surely bounded.
If the square-root of the alpha- ($L^1$-) mixing coefficient
$\phi_1(t)$ is integrable then $\tilde{V}_z$ has  finite second
moment.  \end{itemize}
\end{proposition}

\section{Proof of Lemma~2}
(i) Estimation of $\sup_{|z|\leq z_0\atop |\by|\leq L}
\IE\lt[\lt(\delta_\gamma {V}^\ep_{ z}\rt)^2\rt] (\by):$
 We have that for $\gamma \rho\leq 1$
\beqn
\lefteqn{\sup_{|z|\leq z_0}
\IE\lt[\lt(\delta_\gamma {V}^\ep_{
z}(\bx, \by)\rt)^2\rt]}\\
&=&\sup_{|z|\leq z_0}\int 
4|\sin{(\gamma\by\cdot\bk/2)}|^2
\Phi_{(\eta,\rho)}(\xi,\bk)d\xi d\bk\\
&\leq & 
\sup_{|z|\leq z_0} \int
\lt|\gamma\by\cdot\bk\rt|^2
\Phi_{(\eta,\rho)}(\xi,\bk)d\xi d\bk\\
&\leq & c_9 \gamma^2|\by|^2\sup_{|z|\leq
z_0} 
 \int_{|\kvec|\leq \rho}
(\eta^2+|\bk|^2+|\xi|^2)^{-H-(d+1)/2} |\bk|^{d+1} d\xi
d|\bk|\\
&\leq & c_{10} \gamma^2|\by|^2\sup_{|z|\leq
z_0}  \int_{|\xi|\leq
\rho} 
\int_{|\bk|\leq
\rho}
(\eta^2+|\bk|^2+|\xi|^2)^{-H-(d+1)/2} |\bk|^{d+1}d|\bk| 
d\xi\\ 
&\leq& c_{11} \gamma^2|\by|^2 \sup_{|z|\leq
z_0}\int_{|\xi|\leq \rho}
(\eta^2+|\xi|^2)^{-H+1/2}d\xi\\ &\leq &
 c_{12} \gamma^2|\by|^2
\int_{|\xi| \in (\eta,\rho)}
|\xi|^{-2H+1}d\xi\\ &\leq &
c_{13}\gamma^2|\by|^2
\lt( \eta^{2-2H}
+\rho^{2-2H}\rt).
\eeqn 
For $\rho\gamma\geq 1$  we divide the domain of
integration into $I_0=\{|\bk|\leq \gamma^{-1}\}$ and
$I_1=\{|\bk|\geq \gamma^{-1}\}$ and estimate
their contributions separately. For $I_0$ the upper bound
is similar to the above, namely,
we have
\beqn
\sup_{|z|\leq z_0}\int_{I_0}4
|\sin{(\gamma\by\cdot\bk/2)}|^2
\Phi_{(\eta,\rho)}(\xi,\bk)d\xi d\bk&\leq &
c_{13}\gamma^2|\by|^2
\lt( \eta^{2-2H}
+\gamma^{-2+2H}\rt).
\eeqn
For $I_1$ we have instead that
\beqn
&&\sup_{|z|\leq z_0}\int_{I_1}4
|\sin{(\gamma\by\cdot\bk/2)}|^2
\Phi_{(\eta,\rho)}(\xi,\bk)d\xi d\bk\\
&\leq & 4
\sup_{|z|\leq z_0} \int_{I_1}
\Phi_{(\eta,\rho)}(\xi,\bk)d\xi d\bk\\
&\leq & c_{14} \sup_{|z|\leq
z_0}  \int_{|\xi|\leq
\rho} 
\int_{|\bk|\leq
\rho}
(\eta^2+|\bk|^2+|\xi|^2)^{-H-(d+1)/2} |\bk|^{d-1}d|\bk| 
d\xi\\ 
&\leq& c_{15} \sup_{|z|\leq
z_0}\int_{|\xi|\leq \rho}
(\eta^2+|\xi|^2)^{-H-1/2}d\xi\\ &\leq &
 c_{16}
\int_{|\xi| \in (\gamma^{-1},\rho)}
|\xi|^{-2H-1}d\xi\\ &\leq &
c_{17}
\lt( \gamma^{2H}
+\rho^{-2H}\rt).
\eeqn
Put together, the upper bound becomes
\beqn
{\sup_{|z|\leq z_0\atop |\by|\leq L}
\IE\lt[\lt(\delta_\gamma {V}^\ep_{
z}(\bx,
\by)\rt)^2\rt]}&\leq&\tilde{C}\gamma^2
\lt|\min{(\gamma^{-1},\rho)}\rt|^{2-2H},
\quad 
\gamma,\eta\leq 1\leq
\rho.
\eeqn

(ii) Estimation of $
\sup_{|z|\leq z_0} \IE\lt[\tilde{V}^\ep_{z} (\bx)\rt]^2$:
It follows from the argument for Corollary~1 that
\beqn
\IE\lt[\tilde{V}^\ep_{z} (\bx)\rt]^2
&\leq & \lt(\int^\infty_0 \corr(t) dt\rt)^2
\IE[V^\ep_z]^2\\
&\leq& c\eta^{-2}\lt(\int^\infty_0 r_{1,\rho}(t)dt\rt)^2
\eta^{-2H}.
\eeqn

(iii) Estimation of 
$\sup_{|z|\leq z_0 \atop |\by|\leq L}
\IE\lt[\lt(\delta_\gamma\tilde{V}^\ep_{ z}\rt)^2\rt]
(\by)$:
First note that the correlation coefficient for
$\delta_\gamma \tilde{V}^\ep_z$ is bounded from
above by $c r(t)$ for some constant $c>0$. Then we
have as above that
\beqn
\IE\lt[\delta_\gamma\tilde{V}^\ep_{z} (\bx)\rt]^2
&\leq & c_1\lt(\int^\infty_0 \corr(t) dt\rt)^2
\IE[\delta_\gamma V^\ep_z]^2\\
&\leq& c_2\eta^{-2}\lt(\int^\infty_0 r_{1,\rho}(t)dt\rt)^2
\gamma^2
\lt|\min{(\gamma^{-1},\rho)}\rt|^{2-2H}.
\eeqn

(iv) Estimation of $
\sup_{|z|\leq z_0\atop |\by|\leq L}\lt|\nabla_\by
\IE\lt[\delta_\gamma
\tilde{V}^\ep_z\rt]^2(\by)\rt|$:  By the Cauchy-Schwartz
inequality and the preceding calculation we have
\beqn
\lefteqn{\sup_{|z|\leq z_0\atop |\by|\leq L}\lt|\nabla_\by
\IE\lt[\delta_\gamma
\tilde{V}^\ep_z\rt]^2(\by)\rt|}\\
&\leq&c_1
\sqrt{\gamma\IE\lt[\nabla_\bx\tilde{V}^\ep(\bx+\gamma\by/2)+
\nabla_\bx\tilde{V}^\ep(\bx-\gamma\by/2)\rt]^2}
\sqrt{\IE\lt[\delta_\gamma\tilde{V}^\ep(\bx,\by)\rt]^2}\\
&\leq&c_3\lt(\int^\infty_0 \corr(t) dt\rt)^2 \gamma
\IE^{1/2}\lt[\nabla_\bx{V}^\ep\rt]^2
\IE^{1/2}\lt[\delta_\gamma V^\ep_z(\bx,\by)\rt]^2\\
&\leq& c_4\eta^{-2} \lt(\int^\infty_0
r_{1,\rho}(t)dt\rt)^2
\gamma^2
\rho^{1-H}\minrg^{1-H}
\eeqn

(v) Estimation of $
  \sup_{|z|\leq
z_0}\IE\|\pdgx(\cvtil\theta)\|_2^2
$: A similar line of reasoning and 
a straightforward
spectral calculation yield that
\beqn
\IE\|\pdgx(\cvtil\theta)\|_2^2
&=&\IE\|\nabla_\by\cdot\nabla_\bx
\gamma^{-1}\delta_\gamma \tilde{V}^\ep_z \invf\theta\|_2^2\\
&\leq&c_1\IE\|\nabla_\bx^2
\tilde{V}^\ep_z\invf\theta\|_2^2\\
&\leq& c_2 \eta^{-2}\lt(\int^\infty_0
r_{1,\rho}(t)d t\rt)^2\IE\lt[\nabla_\bx^2
V^\ep_z\rt]^2\\
&\leq& c_3\eta^{-2}\lt(\int^\infty_0
r_{1,\rho}(t)d t\rt)^2\rho^{4-2H}.
\eeqn

\section{Proof of Corollary~\ref{cor3}}
By the Cauchy-Schwartz inequality we have the following
calculation:
\beq
\nn
\lefteqn{\IE\lt[\|\cv\theta(\bx,\bp)\cvtil\theta(\by,\bq)\|_2^2\rt]}\\
\nn&\leq& C_1 \lt\{
\lt\|\IE\lt[\cv\theta(\bx,\bp)\cvtil\theta(\by,\bq)\rt]\rt\|^2_2+
\IE\lt[\lt\|\cv\theta(\bx,\bp)\rt\|^2_2\rt]
\IE\lt[\lt\|\cvtil\theta(\by,\bq)\rt\|^2_2\rt]\rt\}\\
&=&C_1\gamma^{-4} \lt\{\lt\|\IE\lt[\delta_\gamma V^\ep_z(\bx,\bx')
\delta_\gamma \tilde{V}^\ep_z(\by,\by')\rt]\cF_2^{-1} \theta(\bx,\bx')
\cF^{-1}_2\theta(\by,\by')\rt\|_2^2\nn\right.\\
&&\quad+\lt.\lt\|\IE\lt[\lt|\delta_\gamma
V^\ep_z\rt|^2\rt]\cF_2^{-1}
\theta\rt\|^2_2\lt\|\IE\lt[\lt|\delta_\gamma
\tilde{V}^\ep_z\rt|^2\rt]\cF_2^{-1}
\theta\rt\|^2_2\rt\}\nn\\
&=&O\lt(\sup_{|\by|\leq L}\IE\lt|\delta_\gamma
V^\ep_z\rt|^2(\by)\IE\lt|\delta_\gamma
\tilde{V}^\ep_z\rt|^2(\by)\rt)\nn
\nn
\eeq
and 
\beq
\nn
\lefteqn{\IE\lt[\|\cv\cvtil\theta\|_2^2\rt]}\\
\nn&\leq &
C_1'\lt\{\gamma^{-4}\int\IE\lt[\delta_\gamma
V^\ep_z\rt]^2\IE\lt[\delta_\gamma
\tilde{V^\ep_z}\rt]^2 \lt(\cF^{-1}_2\theta\rt)^2d\bx d\by
+\lt\|\IE\lt[\cv\cvtil\theta(\bx,\bp)\rt]\rt\|^2_2\rt\}\\
\nn &=& C_1'\gamma^{-4}\lt\{ \int\IE\lt[\delta_\gamma
V^\ep_z\rt]^2\IE\lt[\delta_\gamma
\tilde{V^\ep_z}\rt]^2\lt(\cF^{-1}_2\theta \rt)^2 d\bx d\by
+\lt\|\IE\lt[\delta_\gamma V^\ep_z \delta_\gamma
\tilde{V}^\ep_z\rt]\cF^{-1}_2
\theta(\bx,\by)\rt\|^2_2
\rt\}\\
&=&O\lt(\sup_{|\by|\leq L}\IE\lt|\delta_\gamma
V^\ep_z\rt|^2(\by)\IE\lt|\delta_\gamma
\tilde{V}^\ep_z\rt|^2(\by)\rt)\nn
\eeq
and
\beq
\nn
\lefteqn{\IE\lt[\|\cvtil\cvtil\theta\|_2^2\rt]}\\
\nn&\leq &
C_2\lt\{\gamma^{-4}
\int\IE\lt[\delta_\gamma
\tilde{V}^\ep_z\rt]^2\IE\lt[\delta_\gamma
\tilde{V}^\ep_z\rt]^2\lt(\cF^{-1}_2\theta \rt)^2 d\bx d\by
+\lt\|\IE\lt[\cvtil\cvtil\theta(\bx,\bp)\rt]\rt\|^2_2\rt\}\\
\nn &=& C_2\gamma^{-4}\lt\{ \int \lt(\IE\lt[\delta_\gamma
\tilde{V}^\ep_z\rt]^2\rt)^2\lt(\cF^{-1}_2\theta \rt)^2 d\bx d\by
+\lt\|\IE\lt[\delta_\gamma \tilde{V}^\ep_z \delta_\gamma
\tilde{V}^\ep_z\rt]\cF^{-1}_2
\theta(\bx,\by)\rt\|^2_2
\rt\}\\
&=&O\lt(\sup_{|\by|\leq L}\IE^2\lt|\delta_\gamma
\tilde{V}^\ep_z\rt|^2(\by)\rt)\nn
\nn
\eeq
where $C_1, C_1', C_2$ are constants independent of $\rho,\eta,
\gamma$ and $L$ is the radius of the ball containing the support
of $\cF^{-1}_2\theta$.  Similarly
we have that
\beqn
\IE \lt\|\cv\cvtil\cvtil\theta\rt\|^2_2
=O\lt(\sup_{|\by|\leq L}\IE^2\lt|\delta_\gamma
\tilde{V}^\ep_z\rt|^2\IE\lt|\delta_\gamma
V^\ep_z\rt|^2\rt).
\eeqn

\end{appendix}

\commentout{
 By the
assumption of the sixth order quasi-scale-invariance, the above terms can be
bounded by, up to a constant factor
\beq
\lt\|\IE\lt[\cv\theta(\bx,\bp)\cvtil\theta(\by,\bq)\rt]\rt\|^2_2+
\IE\lt[\lt\|\cv\theta(\bx,\bp)\rt\|^2_2\rt]
\IE\lt[\lt\|\cvtil\theta(\by,\bq)\rt\|^2_2\rt]
\eeq
and
\beq
\IE\lt[\lt\|\cv\theta(\bx,\bp)\rt\|^2\rt]\IE\lt[\lt\|\cvtil\theta(\bx,\bp)\rt\|^2\rt]
+\lt\|\IE\lt[\cv\cvtil\theta(\bx,\bp)\rt]\rt\|^2_2
\eeq
respectively.
In view of  (\ref{19})  we have the following remain square-integrable uniformly
with respect to
$\ep$.
}

\end{document}